\documentclass{ws-ijmpa}
\usepackage[super,compress]{cite}
\usepackage{graphicx}
\usepackage{amssymb}
\newcommand{\beq}{\begin{eqnarray}}
\newcommand{\eeq}{\end{eqnarray}}
\newcommand{\ba}{\begin{array}}
\newcommand{\ea}{\end{array}}
\newcommand{\nn}{\nonumber}
\newcommand{\no}{\nonumber\\}
\begin{document}
\markboth{Pietro Longhi, Roberto Soldati}
{Neutral Massive Spin $\frac12$ Particles Emission in a Rindler Space}
%
%
\catchline{}{}{}{}{}
%
\title{
NEUTRAL MASSIVE SPIN $\frac12$ PARTICLES \\
EMISSION IN A RINDLER SPACE
}
\author{PIETRO LONGHI}
\address{NHETC and Department of Physics and Astronomy\\
Rutgers, The State University of New Jersey\\
136 Frelinghuysen Road,\\
Piscataway, NJ 08854-8019 USA\\
longhi@physics.rutgers.edu}
\author{ROBERTO SOLDATI}
\address{Dipartimento di Fisica, Universit\`a
di Bologna\\ 
Istituto Nazionale di Fisica Nucleare, Sezione di Bologna\\
Via Irnerio 46, 40126 - Bologna (Italy)\\
roberto.soldati@bo.infn.it}

\maketitle

\begin{history}
\received{Day Month Year}
\revised{Day Month Year}
\end{history}

\begin{abstract}
The Unruh effect for the rate of emission and absorption of neutral massive Majorana spinor particles
-- plausible consituents of the Dark Matter -- 
%
%
in a Rindler space-time is thoroughly investigated. The corresponding Bogolyubov coefficients are
explicitly calculated and the consistency with Fermi-Dirac statistics and the Pauli principle is
actually verified.

\end{abstract}
\ccode{03.70.+k,98.80.Qc}
\def\dirac{i\partial\!\!\!/}
\def\Dirac{iD\!\!\!\!/}
\def\P{P\!\!\!\!/}
\def\DDirac{i\partial\!\!\!/+eA\!\!\!/}
\def\EDirac{D\!\!\!\!/\,}
\def\e{{\rm e}}
\def\egamma{\buildrel-\over\gamma}
\def\eE{\buildrel-\!-\over E}
\def\ex{\buildrel-\over x}
\def\eA{\buildrel-\over A}
\def\oq{\buildrel-\over q}
\def\pa{\partial}
\def\parl{\!\buildrel \leftrightarrow\over\pa\!\!}
\def\d{i\,\partial\!\!\!/}
\def\sparl{\!\buildrel \leftrightarrow\!\!\!\!\over\d}
\def\lg{\left\lgroup}
\def\rg{\right\rgroup}
\def\rmd{{\rm d}}
\def\rmi{{\rm i}}
\def\vu{\upsilon}
\def\ve{\varepsilon}
\def\sumint{\sum\!\!\!\!\!\!\!\!\int}
\def\lbar{^-\!\!\!\!\lambda}
\def\R{\Re{\rm e}}
\def\I{\Im{\rm m}}\
\def\fl{}
\def\ts{\textstyle}
\def\x{\;\!}
\def\rem{$\diamond$}
\section{Introduction}
There is little doubt that the very existence, the nature as well as the eventual detection
of non-baryonic Dark Matter is the most intriguing issue of modern Cosmology and Particle Physics.
There is no surprise that, in spite of the fact that its particle structure is insofar unknown --
certainly rather different from any other kind of ordinary Visible Matter -- 
the nearly exhaustive list of the existing
references\cite{review} attempts to frame Dark Matter within the nearest
extensions of the Standard Model of (visible) 
Particle Physics or
its super-symmetric extensions. In our opinion, this might be a somewhat conventional wisdom and
widespread prejudice which, strictly speaking, does not lie neither upon any solid ground nor any
actual phenomenological evidence. Nonetheless, the non-baryonic Dark Matter is usually assumed
to be a weakly interacting massive particle (WIMP) yet undiscovered. Moreover, the Dark Matter is
also assumed to be stable on the scale of the cosmological structures formation.
More or less well motivated Particle Physics candidates have been proposed\cite{review}, 
all of which arise
from specific models beyond the Standard Model of Particle Physics.
Certainly, if the Dark Matter really exists, the massive, neutral, spin $ \frac12 $
Majorana particles are pretty available candidates for its constituents. Moreover, it is now
quite evident \cite{perlmutter,riess} that galaxies do (uniformly)
accelerate one respect to each other.
Thus it appears to be quite plausible that the eventual detection of the Dark Matter constituents
will be shaped like a (very weakly) accelerated phenomenon. Accelerated observers are expected to
experience the so-called Unruh effect \cite{Unruh,crispino}. This entails that, for instance,
a cosmological thermal bath of WIMP particles is expected to be produced by the cosmic
acceleration, with a characteristic temperature called the Unruh temperature 
$ T={\hbar H}/{2\pi k_{\x\rm B}} $, where 
$ k_{\x\rm B} $ is the Boltzmann's constant and $ H $ the (time dependent) Hubble's parameter, 
which is tightly related to the cosmic acceleration
$ \mathrm a_{\x\rm cosmic} = c H\;\approx\;2.1\times10^{\x-\x9}\ \rm m\ s^{\x-2}$.
In other words, it means that for the uniformly accelerated observers co-moving with the galaxies,
the environment is populated by some thermal distribution of particles created by
the underlying "inertial vacuum" where the relativistic quantum laws of Physics are presumably
holding true. If this is the picture, the corresponding WIMP production is determined by the
so-called Bogolyubov coefficients \cite{NNB,BD} for the inertial to accelerated observers transformation law.

The Bogolyubov coefficients for the production of spinless particles out of the inertial
vacuum have been known for a long time \cite{Unruh} and have been recently rigorously reobtained
\cite{LS}. It is the main purpose of this paper to calculate the Bogolyubov coefficients for the emission
out of the inertial vacuum of a thermal distribution of neutral spin $ \frac12 $ massive particles,
i.e. the most plausible WIMP candidates for the Dark Matter structure. As a matter of fact, to the
best of our knowledge the Fulling modes expansion \cite{fulling} and the ensuing Bogolyubov
coefficients have never been explicitly calculated for the important case of a Majorana spinor field
-- see e.g. the recent up-to-date review paper \cite{crispino}. 
It is just the aim of this paper to fill this gap.\\
%
%
Another purpose of the present paper is to present a first-principles and direct calculation of the effect: in the light of recent controversies regarding the very existence of the Unruh effect, raised by \cite{fedotov} et seq., we aim for a detailed and exhaustive derivation.

%
%
In Section \ref{sec:Majorana-Rindler} we carefully study the Majorana field in a Rindler spacetime. 
In particular we obtain the explicit solutions to the field equation in the accelerated frame (also known as 
\emph{Fulling modes}), we then perform explicit checks of orthonormality and completeness. 
With such solutions at hand, and upon checking self-adjointness of the Majorana Hamiltonian in the 
accelerated frame, we proceed with canonical quantization, establishing the Fock space for the Rindler 
observer.\\
Section \ref{sec:Bogolyubov} contains the explicit calculation of the Bogolyubov coefficients for the massive 
Majorana field in a Rindler space-time. As a crucial and nontrivial check, we show that the Bogolyubov 
coefficients indeed agree with the Pauli exclusion principle, hence providing a consistency check for 
the canonical anticommutation relations adopted in the Rindler space-time. 
A short discussion is contained in the concluding Section.

\section{Majorana Equation in Rindler Coordinates} \label{sec:Majorana-Rindler}
%
%
This section is devoted to the detailed study of the generally covariant Majorana equation 
in a Rindler space-time and of its solutions. Our conventions on tetrads and the spin-connection are summarized in \ref{sec:tetrads}.

\subsection{A short review of Rindler geometry}
Consider the four dimensional Minkowski space with the line element \cite{BD}
\beq
\rmd s^2=\eta_{\,\alpha\beta}\,\rmd X^{\alpha}\rmd X^{\beta}
=g_{\,\mu\nu}(x)\,\rmd \mathrm x^{\,\mu}\rmd \mathrm x^{\,\nu}
\eeq
where the constant metric tensor $\eta_{\,\alpha\beta}={\rm diag}\,(+,-,-,-)$
is relative to an inertial coordinate system in the Minkowski space
labelled by the so called anholomic indices denoted with the greek letters 
from the first part of the alphabet, while the
Einstein convention on the sum over repeated indices is understood.
We employ natural units $\hbar=c=1$ unless explicitly
stated. If we set
\beq
X^{\alpha}=\left(\,c\x\tau,X,Y,Z\,\right)\qquad\quad
\mathrm x^{\,\mu}=(c\x t,x,y,z)
\eeq
Then we shall denote the following 
space-like region of the Minkowski space, viz.,
\beq
{\mathfrak W}_R = \{X^\mu\in{\mathbb R}^4\,|\,X\ge0\,,\,c^2\tau^2\le X^2\}
\eeq 
as the right Rindler wedge.
Here we introduce the so called Rindler curvilinear coordinate system,
which describes an accelerated observer: namely,
\beq
c\x\tau=x\sinh({\rm a}\x t/c)\qquad
X=x\cosh({\rm a}\x t/c) \qquad%
(\,{\rm a}>0\,\vee\,x\ge0\,) %
\label{RRW01}\\
Y=y\qquad\qquad\qquad\ \ \ Z=z \ \qquad\qquad\qquad\qquad\qquad\qquad\qquad
\eeq
where
$[\,{\rm a}\,]$ is the constant acceleration.
The above coordinate transformations
can be readily inverted, viz.,
\beq
\qquad
t=\frac{c}{\rm a}\,{\rm Arth}\,\frac{c\x\tau}{X}\qquad
x=\sqrt{X^2-c^{\x2}\x\tau^2}\ge0\qquad y=Y\qquad z=Z
\eeq
in such a manner that we can also write
\beq
\rmd s^2=g_{\,\mu\nu}(x)\,\rmd \mathrm x^{\,\mu}\rmd \mathrm x^{\,\nu}
&=&x^2({\rm a}/c)^2\,\rmd t^2-\rmd x^2-\rmd Y^2-\rmd Z^2
\eeq
whence we obtain
\beq
g_{\,\mu\nu}(x)=
\left\lgroup
\ba{cccc}
{\rm a}^2x^2/c^4 & 0 & 0 & 0\\
0 & -\,1& 0 & 0\\
0 & 0 & -\,1 & 0\\
0 & 0 & 0 & -\,1 
\ea
\right\rgroup
\eeq
so that
\beq
g= {\rm det}\,g_{\mu\nu}(x)=[\,{\rm det}\,g^{\,\mu\nu}(x)\,]^{-1}=\,
-\,{\rm a}^2x^2/c^4
\eeq
Moreover we find, for $\xi=\mathrm a\x x/c^{\x2}$ and $\eta=\mathrm a\x t/c\,,$
\begin{equation}\label{eq:coordinate-change}
\frac{\partial X^{\alpha}}{\partial\x\mathrm x^{\x\nu}} =
\left\lgroup
\begin{array}{cccc}
\xi\cosh\eta& \sinh\eta & 0 & 0\\
\xi\sinh\eta & \cosh\eta & 0 & 0\\
0&0&1&0\\
0&0&0&1
 \end{array}
\right\rgroup\equiv\,\zeta^{\x\alpha}_{\x\nu}(\xi,\eta)
\end{equation}
which endorses
\[{\rm det}\x({\partial X^{\alpha}}/{\partial\x\mathrm x^{\x\nu}})=\sqrt{-g}=\xi=\mathrm a\x x/c^{\x2}>0\]
In order to set notation, we denote the inverse of (\ref{eq:coordinate-change}) by
\begin{equation}
 \frac{\partial\x\mathrm x^{\x\nu}}{\partial X^{\alpha}}=\zeta_{\x\alpha}^{\x\nu}[\,\mathrm X(\xi,\eta)\,].
\end{equation}
%
%
Notice that, by changing both signs in the definitions
(\ref{RRW01}), we shall cover
the other space-like region of the Minkowski space,
i.e. the left Rindler wedge
\beq
{\mathfrak W}_L = \left\{X^\alpha\in{\mathbb R}^4\,|\,X\le0\,,\,c^2\tau^2\le X^2\right\}
\eeq

\subsection{The generally covariant Majorana equation}
The Majorana equation, in its generally covariant form, is
\begin{equation}\label{eq:covariant-Majorana}
\qquad
 V_\alpha^\mu(\mathrm x)\,\gamma^{\,\alpha}_{M}\left\{\partial_{\mu} - \mathrm i\x\Gamma_{\mu}(\mathrm x)\right\} \psi_M(\mathrm x)
+im\,\psi^{\,c}_M(\mathrm x)=0
\qquad\quad
\psi_M=\psi^{\,c}_M
\end{equation}
The particular case of the uniformly accelerated noninertial frame referred to the 
Rindler's curvilinear coordinates system can be handled as follows.
As in the previous section, let $\mathrm X$ denote the coordinates in the frame of an inertial observer, and $\mathrm x$ those of of the accelerated observer, then we define
\begin{eqnarray}
\Lambda(t)=\left\lgroup\begin{array}{cccc}
\cosh\eta&\sinh\eta&0&0\\
\sinh\eta&\cosh\eta&0&0\\
0&0&1&0\\
0&0&0&1\end{array}\right\rgroup\qquad\quad\Lambda(0)=\mathbb I\\
 \Psi_M[\x\mathrm X(\mathrm x)\x]= D[\x\Lambda(t)\x]\,
\psi_M(\mathrm x)
\label{spinortransform}\\
D[\x\Lambda(t)\x]=
\exp\left\lbrace -\,\mathrm i\,\sigma_{M}^{\x01}\,\omega_{01}(t)\right\rbrace
=\exp\left\lbrace\textstyle\frac12\,\alpha_{M}^{\x1}\,\eta\right\rbrace\nonumber\\
=\cosh\textstyle\frac12\eta
+ \alpha_{M}^{\x1}\,\sinh\frac12\eta
= D^{\x\dagger}[\x\Lambda(t)\x]
\end{eqnarray}
together with
\beq\label{eq:vb}
&& \zeta^{\x\alpha}_{\x\mu}(\eta,\xi)=\Lambda^{\alpha}_{\beta}(t)\,{V}^{\,\beta}_{\mu}(x) \\
&& {V}^{\,\mu}_{0}(x)=\xi^{\x-1}\,{\delta^{\x\mu}_{\x0}}\;
\qquad
{V}^{\,\mu}_{\jmath}=\delta^{\x\mu}_{\x\jmath}
\eeq
where the one-parameter rank-four square matrix $D[\x\Lambda(t)\x]$ represents a local boost
along the $OX-$direction.
Then it is clear that the only contribution to (\ref{eq:covariant-Majorana})
%
%
comes from the term
\begin{eqnarray*}
 \frac{\partial}{\partial t}\left\lbrace 
D[\x\Lambda(t)\x]\,\psi_M(\mathrm x)\right\rbrace
&=& D[\x\Lambda(t)\x]\,\left\lbrace \frac{\partial}{\partial t}
-\mathrm i\x\Gamma_0\right\rbrace \Psi_M(\mathrm x)
\end{eqnarray*}
\[
\Gamma_0 = \ts\frac12\,i{\rm a}\,\gamma_M^0\gamma_M^1
=\ts\frac12\,i{\rm a}\,\alpha_M^1
=\ts\frac12\,i{\rm a}\left\lgroup\ba{cc}
-\,\sigma_1 & 0\\
0 & \sigma_1\ea\right\rgroup
\]
Thus, in this specific case we can write
\begin{eqnarray*}
&& \Lambda_{\alpha}^{\beta}(t)\,V^{0}_{\beta}(x)\,
D^{-1}[\,\Lambda(t)\,]\,\gamma_M^\alpha
\left\lbrace {\mathrm i\x\partial_{t}}D[\,\Lambda(t)\,]\right\rbrace\\
&=&\Lambda_{\alpha}^{\beta}(t)\,V^{0}_{\beta}(x)\,
D^{-1}[\,\Lambda(t)\,]\,\gamma_M^\alpha\,D[\x\Lambda(t)\x]\,\Gamma_0\\
&=&\Lambda_{\alpha}^{\beta}(t)\,V^{0}_{\beta}(x)\,\Lambda^{\alpha}_{\x\eta}(t)\,
\gamma^{\x\eta}_{M}\,\Gamma_0=V^{0}_{\beta}(x)\gamma^{\x\beta}_{M}\,\Gamma_0\\
&=&\frac{c^{\x2}}{\mathrm a\x x}\;\gamma^{\x0}_{M}\,\Gamma_0
\end{eqnarray*}
in such a manner that we eventually obtain
\begin{eqnarray*}
\left\lbrace 
 V_0^0(x)\gamma_M^0\left( {\partial}_{0} - \mathrm i\,\Gamma_{0}\right)
+ \gamma_M^{\x k}\,\partial_{k}
+ im\right\rbrace \psi_M(\mathrm x) = 0\qquad\quad
\psi_M=\psi_M^{\x c}
\end{eqnarray*}

\medskip
It is worthwile to rederive this result in a manifestly generally covariant formalism.%
Turning to a general
curved space and to some generic curvilinear coordinate system $\mathrm x^{\x\mu}=(ct,x,y,z)$,
then we come \cite{BD,weinberg} to the generally covariant Majorana bispinor wave equation
\beq
V_\alpha^\mu(\mathrm x)\,\gamma^{\,\alpha}_{M}\,\nabla_\mu\psi_M(\mathrm x)
+im\,\psi^{\,c}_M(\mathrm x)=0
\qquad\quad
\psi_M=\psi^{\,c}_M
\label{covdiracequ}
\eeq
where
\beq
\nabla_\mu\equiv\pa_\mu - \mathrm i\Gamma_\mu=\pa_\mu 
- {\textstyle\frac12}\,\mathrm i\,\sigma^{\,\alpha\beta}_{M}\,\omega_{\,\alpha\beta\,;\,\mu}\\
\sigma^{\,\alpha\beta}_{M}\,\equiv\,{\textstyle\frac14}\,
\mathrm i\,[\,\gamma^{\,\alpha}_{M}\,,\,\gamma^{\,\beta}_{M}\,]\\
\omega_{\,\alpha\beta\,;\,\mu}\,\equiv\,V_\alpha^\nu(\mathrm x)\,V_{\beta\nu\,;\,\mu}(\mathrm x)\\
V_{\beta\nu\,;\,\mu}(\mathrm x)\equiv D_\mu\,V_{\beta\nu}(\mathrm x)
=\pa_\mu\,V_{\beta\nu}(\mathrm x)-\Gamma_{\mu\nu}^{\lambda}(\mathrm x)\,V_{\beta\lambda}(\mathrm x)
\eeq
%
%
Again, if we turn to the purely imaginary Majorana representation for the
gamma matrices $\gamma_M^{\,\alpha}+\gamma_M^{\,\alpha\x\ast}=0\,,$ 
then the above covariant bispinor equation (\ref{covdiracequ}) admits real self-conjugated
Majorana bispinor solutions
\beq
V_\alpha^\mu(\mathrm x)\,\gamma_M^{\,\alpha}\,\nabla_\mu\,\psi_M(\mathrm x) + im\,\psi^\ast_M(\mathrm x)=0
\qquad\quad\psi_M=\psi^{\,c}_M=\psi_M^\ast
\label{covmajequ}
\eeq
It turns out that,
taking into account the simplest choice (\ref{eq:vb}) for the {\it vierbeine} or tetrad fields,
the matrix-valued spin connection components are given by
\beq
\Gamma_0(x)&=&{\textstyle\frac i4}\,\gamma_M^0\gamma_M^1\,\left[\,V_0^\nu(x)\,V_{1\nu\,;\,0}(x)
- V_1^\nu(x)\,V_{0\nu\,;\,0}(x)\,\right]\\
\Gamma_1(x)&=&{\textstyle\frac i4}\,\gamma_M^0\gamma_M^1\,\left[\,V_0^\nu(x)\,V_{1\nu\,;\,1}(x)
- V_1^\nu(x)\,V_{0\nu\,;\,1}(x)\,\right]\\
&&\Gamma_2(x)=\Gamma_3(x)=0
\eeq
Now we have
\beq
V_0^\nu\,V_{1\nu\,;\,0}= -\,V_0^\nu\,\Gamma_{0\nu}^\kappa\,V_{1\kappa}
=\rm a\\
V_1^\nu(x)\,V_{0\nu\,;\,0}(x) = -\,V_1^1\,\Gamma_{10}^{\,0}\,V_{00}=-\,{\rm a}
\eeq
that yields once again as before
\beq
\Gamma_0=\ts\frac12\,i{\rm a}\,\gamma_M^0\gamma_M^1
=\ts\frac12\,i{\rm a}\,\alpha_M^1
=\ts\frac12\,i{\rm a}\left\lgroup\ba{cc}
-\,\sigma_1 & 0\\
0 & \sigma_1\ea\right\rgroup
\eeq
Finally one gets
\beq
\Gamma_1(x)\equiv0
\eeq
in such a manner that we can eventually write
\beq
\qquad
\nabla_0=\pa_t + \ts\frac12\,{\rm a}\,\alpha_M^1
\qquad\quad
\nabla_\jmath=\pa_\jmath\qquad\quad(\,\jmath=x,y,z\,)
\eeq

Hence, for a uniformly accelerated noninertial observer employing a Rindler's
curvilinear coordinate system, the covariant Majorana massive operator definitely reads
%
%
\beq
iD\!\!\!\!/_M-m&\equiv& iV_\alpha^\mu(x)\,\gamma_M^{\,\alpha}\,\nabla_\mu -m\no
&=&V_0^0(x)\gamma_M^0\left(i\pa_0+\Gamma_0\right)+\gamma_M^{\,k}\,i\pa_k-m \no
&=& \frac{1}{{\rm a}x}\,\gamma_M^0\,i\pa_t
-\gamma_M^1\,\hat P_x -\gamma_M^2\,\hat p_y - \gamma_M^3\,\hat p_z - m
\eeq
in which we have set
\[
\hat P_x
=\hat p_x - \frac{i\hbar}{2x}=-\,i\hbar\left(\frac{\pa}{\pa x}+\frac{1}{2x}\right)\qquad\quad
\hat p_\jmath = -\,i\hbar\x\pa_\jmath\qquad\quad
(\;\jmath=x,y,z\,)
\]
\subsection{Self-adjointness of the self-conjugated Hamiltonian}
On the one hand, from the above explicit expression of the covariant Dirac
operator in the Majorana representation of the gamma matrices,
it keeps manifestly true that even the covariant Majorana
spinor equation (\ref{covmajequ}) always admits real solutions, i.e.
the self-conjugated spinors, as expected
from the general covariance and the equivalence principles.
On the other hand, the Rindler time evolution of a real Majorana bispinor
is governed by the 1-particle Majorana Hamiltonian 
\[
i\hbar\x\pa_t\x\psi_M(t,{\bf x})\;=\;H_M\x\psi_M(t,{\bf x})
\]
which can be readily extracted from the covariant equation (\ref{covmajequ})
and reads
\beq 
\qquad\quad
H_M=\sqrt{-g}\,\left\{\alpha^1_M\left(\hat p_x - \frac{i}{2x}\right)
+\alpha^2_M\,\hat p_y+\alpha^3_M\,\hat p_z + {m}\,\beta_M\right\}
=H_M^{\x\dagger}
\label{majham}
\eeq
where, as customary,
\beq
\qquad
\alpha^{\,k}_M=\gamma^0_M\gamma^k_M\qquad\quad\hat p_k=
-\,i\hbar\pa_k\qquad\quad\beta_M=\gamma^0_M
\qquad\quad(\,k=1,2,3\,)
\eeq
Just like in the Dirac case, the 1-particle
Majorana Hamiltonian is self-adjoint
in the right Rindler wedge $\mathfrak W_R$ iff
\beq
\lim_{\lambda\,\to\,1^+}\;\lim_{x\,\to\,0^+}
\int\rmd^2{\bf x}_\perp\;x^{\x\lambda}\,\psi_{\x2}^{\x\dagger}(t,x,{\bf x}_\perp)\,
\alpha^1_M\,\psi_1(t,x,{\bf x}_\perp)=0\qquad
\forall\,\psi_1\,,\,\psi_{\x2}\in\mathfrak D_M
\eeq
the order of the two limits being not interchangeable.

\medskip
As a matter of fact the Rindler time evolution of a complex Dirac bispinor
is governed by the 1-particle Dirac Hamiltonian 
\[
i\hbar\x\pa_t\x\psi(t,{\bf x})\;=\;H_D\x\psi(t,{\bf x})
\]
which can be readily extracted from the covariant equation (\ref{covmajequ})
and reads
\beq 
H_D=\sqrt{-g}\,\left\{\alpha^1\left(\hat p_x - \frac{i}{2x}\right)
+\alpha^2\,\hat p_y+\alpha^3\,\hat p_z + {m}\,\beta\right\}
\eeq
where, as customary,
\beq
\qquad
\alpha^{\,k}=\gamma^0\gamma^k\qquad\quad\hat p_k=
-\,i\hbar\pa_k\qquad\quad\beta=\gamma^0
\qquad\quad(\,k=1,2,3\,)
\eeq
Let us investigate a little bit closer the hermiticity property
of the above Dirac Hamiltonian operator $H_D\,.$
The general diffeomorphisms invariant inner product 
between any arbitrary pair of square integrable complex solutions of the covariant Dirac equation
is defined by
\beq
(\,\psi_{\x2}\,,\,\psi_1\,)\equiv\oint_\Sigma\,V_{\alpha\mu}(x)\,\bar\psi_{\x2}(\mathrm x)\,
\gamma^{\,\alpha}\,\psi_1(\mathrm x)\,\rmd\Sigma^{\,\mu}
\eeq
where $\bar\psi(\mathrm x)\equiv\psi^{\x\dagger}(\mathrm x)\,\gamma^0=\psi^{\x\dagger}(\mathrm x)\,\beta$ 
whereas $\Sigma$ is a three dimensional future oriented space-like hyper-surface.
Again, for the initial time three dimensional hyper-surface 
in the right Rindler wedge
$\mathfrak W_R$ we get
\beq
\rmd\Sigma^{\x0}\,=\,-\;\frac{\theta(x)\x\rmd x}{\surd(-\,g)}\;\rmd^2{\bf x}_\perp
\qquad\rmd\Sigma^{\x\imath}=0\quad(\,\imath=1,2,3\,)
\eeq
and since we have $V_{00}=\sqrt{-\,g}$ we eventually obtain
\beq
(\,\psi_{\x2}\,,\,\psi_1\,)\,=\;-\int_{0^+}^\infty{\rmd x}\int\rmd^2{\bf x}_\perp\;
\psi_{\x2}^{\x\dagger}(t,x,{\bf x}_\perp)
\,\psi_1(t,x,{\bf x}_\perp)
\eeq
Thus we find
\beq
(\,\psi_{\x2}\,,\,H_D\,\psi_1\,)&=&
-\int_{0^+}^\infty{\rmd x}\int\rmd^2{\bf x}_\perp\;
\psi_{\x2}^{\x\dagger}(t,x,{\bf x}_\perp) H_D
\,\psi_1(t,x,{\bf x}_\perp)\no
&=&-{\rm a}\int\rmd^2{\bf x}_\perp\int_{0^+}^\infty x\,\rmd x\;
\psi_{\x2}^{\x\dagger}(t,x,{\bf x}_\perp)\no
&\times&\left\{\alpha^1\left(\hat p_x - \frac{i\hbar}{2x}\right)
+\alpha^2\,\hat p_y+\alpha^3\,\hat p_z + {m}\,\beta
\right\}\,\psi_1(t,x,{\bf x}_\perp)\nn
\eeq
Let us focus our attention on the expression
\beq
i\hbar{\rm a}\int_{0^+}^\infty x\;
\psi_{\x2}^{\x\dagger}(t,x,{\bf x}_\perp)\,
\alpha^1\left(\pa_x + \frac{1}{2x}\right)
\psi_1(t,x,{\bf x}_\perp)\,\rmd x\no
=\;i\hbar{\rm a}\left[\,x\,
\psi_{\x2}^{\x\dagger}(t,x,{\bf x}_\perp)\,
\alpha^1\,\psi_1(t,x,{\bf x}_\perp)\,\right]_{0^+}^\infty\no
-\,i\hbar{\rm a}\int_{0^+}^\infty x\;
\left[\,\left(\pa_x + \frac{1}{2x}\right)
\psi_{\x2}(t,x,{\bf x}_\perp)\,\right]^{\x\dagger}
\alpha^1\,\psi_1(t,x,{\bf x}_\perp)\,\rmd x\no
\eeq
in such a manner that we can write
\beq
(\,\psi_{\x2}\,,\,H_D\,\psi_1\,)
&=& i{\rm a}\hbar\int\rmd^2{\bf x}_\perp\;\left[\,x\,
\psi_{\x2}^{\x\dagger}(t,x,{\bf x}_\perp)\,
\alpha^1\,\psi_1(t,x,{\bf x}_\perp)\,\right]_{0^+}^\infty\no
&-&{\rm a}\int_{0^+}^\infty x\,\rmd x\int\rmd^2{\bf x}_\perp\,\left[\,
\alpha^1\left(\hat p_x - \frac{i\hbar}{2x}\right)\psi_{\x2}(t,x,{\bf x}_\perp)\,\right]^{\x\dagger}
\psi_1(t,x,{\bf x}_\perp)\no
&+&\left(\,\sqrt{-\,g}\,\{\alpha^2\,\hat p_y+\alpha^3\,\hat p_z
 + {m}\,\beta\}\,\psi_{\x2}\,,\,\psi_1\,\right)
\eeq
This last equality means that we can identify
\beq
H_D^{\,\dagger}=\sqrt{-\,g}\left\lbrace
\alpha^1\left(\hat p_x - \frac{i\hbar}{2x}\right)
+ \alpha^2\,\hat p_y+\alpha^3\,\hat p_z
 + {m}\,\beta\right\rbrace\,=\,H_D
\eeq
which means that the symmetric Dirac Hamiltonian operator is 
actually self-adjoint on the right Rindler wedge
provided the domain $\mathfrak D$ of the Dirac complex bispinors $\psi_1\,,\,\psi_{\x2}$ is such that
\beq
\lim_{\lambda\,\to\,1^+}\int\rmd^2{\bf x}_\perp\;\left[\,x^{\x\lambda}\,
\psi_{\x2}^{\x\dagger}(t,x,{\bf x}_\perp)\,
\alpha^1\,\psi_1(t,x,{\bf x}_\perp)\,\right]_{0^+}^\infty
=0\qquad\quad
\forall\,\psi_1\,,\,\psi_{\x2}\in\mathfrak D
\eeq
\subsection{Solutions of the generally covariant Majorana wave equation}
Consider the second order differential operator
acting on the Majorana spinors, which can be cast into the form
\beq
&-&\left(\,iD\!\!\!\!/_M-m\,\right)\left(\,iD\!\!\!\!/_M+m\,\right)\no
&=&\left(\,V_\alpha^\mu(x)\,\gamma_M^{\,\alpha}\,\,\nabla_\mu + im\,\right)
\left(\,V_\beta^\lambda(x)\,\gamma_M^{\,\beta}\,\,\nabla_\lambda - im\,\right)\no
&=& \frac{1}{{\rm a}^2x^2}\;\pa_t^2 + \hat P_x^2 + \hat p_\perp^2 + m^2
+ \frac{\alpha^1_M}{{\rm a}\,x^2}\,\pa_t\no
&=& \frac{1}{{\rm a}^2x^2}\left(\pa_t^2 + \alpha^1_M\,{\rm a}\,\pa_t\right)
- \left(\,\pa_x+\frac{1}{2x}\,\right)^2
- \pa_\perp^2 + m^2
\eeq
It is convenient to obtain the spinor solutions of the Majorana equation from the
solutions of the second order differential equation
\beq
\left(\,iD\!\!\!\!/_{\,M}-m\,\right)\left(\,iD\!\!\!\!/_{\,M}+m\,\right)\,f(t,x,y,z)\,\Upsilon=0
\label{2ndordmajequ}
\eeq
where $f(t,x,y,z)$ is an invariant scalar function, whereas $\Upsilon$ is a constant
eigenspinor of the matrix 
\[
\alpha^1_M=\gamma^0_M\gamma^1_M=\left\lgroup
\ba{cc}
-\,\sigma_1 & 0\\
0 & \sigma_1
\ea
\right\rgroup
\]
There are two degenerate real eigenvalues $\lambda_{\,\pm}=\pm\,1$ of the Hermitean matrix
$\alpha^1_M$ and for each eigenvalue one pair of degenerate constant eigenbispinors.
Moreover, consider the $OX-$component of the spin in the particle comoving 
instantaneous rest frame,
that is nothing but the helicity Hermitean matrix along the direction of the acceleration
\beq
\textstyle\frac12\,\Sigma_M^1\;\equiv\;\frac{i}{4}\,[\,\gamma_M^2\,,\,\gamma_M^3\,]
\qquad\quad
\Sigma_M^1\;\equiv\;\left\lgroup
\ba{cc}
0 & i\sigma_3\\
-\,i\sigma_3 & 0
\ea\right\rgroup
\eeq
which obviously satisfies $[\,\alpha^1_M\,,\,\Sigma_M^1\,]=0\,,$
in such a manner that we can set
\beq
\Upsilon_{+}^{\,\uparrow}=\textstyle\frac12
\left\lgroup
\ba{c}
1\\ -1\\ -\,i\\ -\,i
\ea\right\rgroup
\quad\qquad
\Upsilon_{+}^{\,\downarrow}=\textstyle\frac12
\left\lgroup
\ba{c}
1\\-1\\i\\i
\ea\right\rgroup
\eeq
\beq
\Upsilon_{-}^{\,\uparrow}=\textstyle\frac12
\left\lgroup
\ba{c}
i\\i\\1\\-1
\ea\right\rgroup
\quad\qquad
\Upsilon_{-}^{\,\downarrow}=\textstyle\frac12
\left\lgroup
\ba{c}
-\,i\\-\,i\\1\\-1
\ea\right\rgroup
\eeq
The above four bispinors do realize a complete and orthonormal set
and fulfill by construction
\beq
\alpha^1_M\,\Upsilon_{\pm}^{\,\uparrow}=\pm\,\Upsilon_{\pm}^{\,\uparrow}
\quad\qquad
\alpha^1_M\,\Upsilon_{\pm}^{\,\downarrow}=\pm\,\Upsilon_{\pm}^{\,\downarrow}
\eeq
\beq
\Sigma_M^1\,\Upsilon_{\pm}^{\,\uparrow}=\Upsilon_{\pm}^{\,\uparrow}
\qquad\qquad
\Sigma_M^1\,\Upsilon_{\pm}^{\,\downarrow}=\,-\,\Upsilon_{\pm}^{\,\downarrow}
\eeq
Notice that if we set
\beq
\chi_{+}=\left\lgroup
\ba{c}
1\\-1
\ea\right\rgroup
\qquad\quad
\chi_{-}=\left\lgroup
\ba{c}
i\\i
\ea\right\rgroup
\eeq
then we can write
\beq
\qquad
\Upsilon_{\pm}^{\,\uparrow}=\textstyle\frac12
\left\lgroup
\ba{c}
\chi_{\pm}\\-\x\sigma_2\chi_{\pm}^{\x\ast}
\ea\right\rgroup
=\Upsilon_{\pm}^{L}
\qquad\quad
\Upsilon_{\pm}^{\,\downarrow}=\textstyle\frac12
\left\lgroup
\ba{c}
\sigma_2\chi_{\pm}^{\x\ast}\\ \chi_{\pm}
\ea\right\rgroup
=\Upsilon_{\pm}^{R}
\eeq
which means that the constant self-conjugated Majorana bispinors $\Upsilon$ of positive helicity
do correspond to the left handed Weyl spinors, while the negative helicity
self-conjugated Majorana bispinors to the right handed Weyl spinors.
Thus we can eventually identify up arrow with left handed $\uparrow\,=L$
and down arrow with right handed $\downarrow\,=R\,.$ 
Moreover we can check by direct inspection that 
\beq
\left(\Upsilon_{\pm}^{L}\right)^*=\Upsilon_{\pm}^{R}
\eeq

There are in general four linearly independent
{\bf complex} solutions of the covariant bispinor equation (\ref{covdiracequ})
which can always be written in the form
\beq
\psi^{\,r}_{\pm}(t,x,y,z)=\left(\,iD\!\!\!\!/_{\,M}+m\,\right)\,
f_\pm(t,x,y,z)\,\Upsilon_{\pm}^{\,r}
\qquad\quad(\,r\,=\,\uparrow\,,\,\downarrow\,=\,L,R\,)
\label{psi_M}
\eeq
but since we have
\beq
\left(\,iD\!\!\!\!/_{\,M}+m\,\right)^\ast
=iD\!\!\!\!/_{\,M}+m
\eeq
thanks to the purely imaginary form of the $\gamma-$matrices
in the Majorana representation, there exists only
two types of real massive bispinor solutions, viz.
\beq
\qquad
\chi^{\,r}_{\pm}(t,{\bf x})=\left(\,iD\!\!\!\!/_{\,M}+m\,\right)\left[\,
f_\pm(t,{\bf x})\,\Upsilon_{\pm}^{\,r}\ +\ {\rm c.c.}\,\right]
\qquad\quad(\,r\,=\,L,R\,)
\label{realmassivebispinor}
\eeq
where $f_\pm(t,x,y,z)$ are arbitrary complex solutions of the second order 
differential equations
\beq
\qquad\left\lbrace
\frac{-1}{{\rm a}^2x^2}\left(\pa_t^{\x2} \pm {\rm a}\,\pa_t\right)
+ \left(\,\pa_x+\frac{1}{2x}\,\right)^2
+ \pa_\perp^{\x2} - m^{\x2}\right\rbrace
f_\pm(t,x,y,z)=0
\eeq
In fact, after taking the partial Fourier transform
\beq
\qquad
f_{\pm}(t,x,{\bf x}_\perp)\equiv\int_{-\infty}^\infty\rmd p_0
\int{\rmd\bf p}\
\tilde f_{\pm}(p_0,{\bf p};x)\,\exp\{-ip_0t+i{\bf p}\cdot\,{\bf x}_\perp\}\\
\qquad\quad
{\bf x}_\perp=(y,z)\qquad\quad {\bf p}=(p_y,p_z)
\qquad\quad\kappa=\sqrt{{\bf p}^2+(mc/\hbar)^2}
\eeq
%
%
and setting 
\beq
ip_0=\bar p\qquad\quad\bar p^2=-\,p_0^2
\eeq
we recover the modified Bessel equations
\beq
\left[\,\frac{\rmd^2}{\rmd x^2}+\frac{1}{x}\cdot\frac{\rmd}{\rmd x}
- \frac{1}{x^2}\left(\frac{\bar p}{\rm a}\mp\frac12\right)^2 -\kappa
\,\right]\tilde f_\pm(\bar p,{\bf p};x)\;=\;0
\eeq
the most general solutions of which are the modified Bessel functions
\beq
\tilde f_\pm(\bar p,{\bf p};x)=
C_\pm\,K_{i\nu\,\mp\frac12}(\kappa x) +
C^{\,\prime}_\pm\,I_{i\nu\,\mp\frac12}(\kappa x)
\quad\qquad\nu=\frac{\bar p}{i\rm a}
\eeq
The solution $I_{i\nu\,\mp\frac12}(\kappa x)$ must be discarded
for $x>0$ so that we are left with the sets
\beq
\qquad
\tilde f_\pm(p_0,{\bf p};x)=\tilde f_{\pm}(\nu,{\bf p})\,K_{i\nu\,\mp\frac12}(\kappa x)
\qquad\quad\nu=\frac{p_0}{\rm a}\qquad (\,x>0\,)
\eeq
and if we take into account that the solutions keep the very same under the substitutions
\[
\tilde f_{\pm}(\nu,{\bf p})\quad\mapsto\quad
\tilde f_{\mp}(-\nu,-{\bf p})\quad\mapsto\quad
\tilde f^{\ast}_{\mp}(\nu,{\bf p})
\]
\[
K_{i\nu\,\mp\frac12}(\kappa x)\quad\mapsto\quad
K_{-i\nu\,\mp\frac12}(\kappa x)=K_{i\nu\,\pm\frac12}(\kappa x)
\]
we come to the two sets of the real bispinor normal modes
on the right Rindler wedge $\mathfrak W_R$
which are solutions of the second order covariant equation
(\ref{2ndordmajequ}), i.e.,
\beq
\phi_{\pm}^{\;\rm p}(t,x,y,z)&=&
\Upsilon_{\pm}^{L}\,
K_{i\nu\,\mp\frac12}(\kappa x)\,
\exp\{-i{\rm a}\x\nu t/c+i{\bf p}\,\cdot\,{\bf x}_\perp\}\no
&+&
\Upsilon_{\pm}^{R}\,K_{i\nu\,\pm\frac12}(\kappa x)\,
\exp\{i{\rm a}\x\nu t/c - i{\bf p}\,\cdot\,{\bf x}_\perp\}
\qquad\quad (\,x>0\,)
\nn
\eeq
where the suffix ${\rm p}$ stands for the triple $(E,{\bf p})$ or $(\nu,{\bf p})$ with $\nu=cE/\hbar\rm a\,.$
The above sets of real bispinor normal modes are eigenstates of the matrix $\alpha_M^1$ because
\[
\alpha_M^1\,\phi_{\pm}^{\;\rm p}(t,x,y,z)=\,\pm\,\phi_{\pm}^{\;\rm p}(t,x,y,z)
\]
\subsection{Majorana bispinor normal modes}
In order to obtain the bispinor solutions (\ref{psi_M})
it is convenient to start with
\beq
\left(\,iD\!\!\!\!/_{\,M}+m\,\right)\,\exp\{-\,i\x E\x t + i\x{\bf p}\,\cdot\,{\bf x}_\perp\}=
\left(\,P\!\!\!\!/_M + m\,\right)\,\exp\{-\,i\x(E\x t - {\bf p}\,\cdot\,{\bf x}_\perp)\}
\eeq
where
\beq
\qquad
P\!\!\!\!/_M + m = \gamma_M^1\,iD_x\,
+ \frac{E}{{\rm a}\x x}\,\gamma_M^0
-\gamma_M^2\,p_y - \gamma_M^3\,p_z + m\;=\no
\left\lgroup\ba{cccc}
D_x+m & -\,i\x(\x p_z-\nu/x\x) & 0 & ip_y\\
-\,i\x(\x p_z + \nu/x\x) & -\,D_x+m & -\,ip_y & 0\\
0 & -\,ip_y & D_x+m & -\,i\x(\x p_z + \nu/x\x)\\
ip_y & 0 & -\,i\x(\x p_z - \nu/x\x) & -\,D_x+m
\ea\right\rgroup
\eeq
Next we find
\beq
\left(\,P\!\!\!\!/_M+m\,\right)\Upsilon_{\pm}^{\,L}\;
=\;m\Upsilon_{\pm}^{\,L} + (\,p_y+ip_z\,)\Upsilon_{\pm}^{\,R}
- \Upsilon_{\mp}^{\,L}\,i\delta_{+}\\
\left(\,P\!\!\!\!/_M+m\,\right)\Upsilon_{\pm}^{\,R}\;
=\;m\Upsilon_{\pm}^{\,R}-(\,p_y-ip_z\,)\Upsilon_{\pm}^{\,L}
+ \Upsilon_{\mp}^{\,R}\,i\delta_{-}
\eeq
in which we have set
\beq
\delta_\pm\;\equiv\;i\hat P_x\pm\,\frac{i\nu}{x}\;=\;
\frac{\rmd}{\rmd x}+\frac{1}{x}\left({\textstyle\frac12}\pm{i\nu}\right)\;
=\;\delta_\mp^{\x\dagger}
\eeq
Now, from the recursion formul\ae
\beq
\delta_\pm\,
K_{i\nu\pm\frac12}(\kappa x)
+\kappa\,K_{i\nu\mp\frac12}(\kappa x)\;=\;0
\eeq
we get
\beq
\qquad
\delta_{\mp}\,\tilde f_\pm(p_0,{\bf p};x)
= \tilde f_\pm(\nu,{\bf p})\,\delta_{\mp}\,K_{i\nu\,\mp\frac12}(\kappa x)
= -\,\kappa\,\tilde f_\pm(\nu,{\bf p})\,K_{i\nu\,\pm\frac12}(\kappa x)
\eeq
and after the suitable introduction of the transverse momentum dependent spin states
\beq
u_{\x\pm}({\bf p})=
m\Upsilon_{\pm}^{L} + (\,p_y+ip_z\,)\Upsilon_{\pm}^{R}\\
v_{\x\pm}({\bf p})=
m\Upsilon_{\pm}^{R}-(\,p_y-ip_z\,)\Upsilon_{\pm}^{L}
\eeq
we can eventually write the two sets of bispinor solutions: namely,
\beq
\left\lbrace\begin{array}{c}
\left(\,P\!\!\!\!/_M+m\,\right)\Upsilon_{+}^{L}\,\tilde f_{+}(p_0,{\bf p};x)=
u_{\x+}({\bf p})\,K_{i\nu\,-\frac12}(\kappa x)
+i\kappa\,\Upsilon_{-}^{L}\,K_{i\nu\,+\frac12}(\kappa x)\\
\left(\,P\!\!\!\!/_M+m\,\right)\Upsilon_{+}^{R}\,\tilde f_{+}(p_0,{\bf p};x)=
v_{+}({\bf p})\,K_{i\nu\,-\frac12}(\kappa x)
- i\kappa\,\Upsilon_{-}^{R}\,K_{i\nu\,+\frac12}(\kappa x)
\end{array}\right.
\eeq
together with
\beq
\left\lbrace\begin{array}{c}
\left(\,P\!\!\!\!/_M+m\,\right)\Upsilon_{-}^{L}\,\tilde f_{-}(p_0,{\bf p};x)=
u_{-}({\bf p})\,K_{i\nu\,+\frac12}(\kappa x)
- i\kappa\,\Upsilon_{+}^{L}\,K_{i\nu\,-\frac12}(\kappa x)\\
\left(\,P\!\!\!\!/_M+m\,\right)\Upsilon_{-}^{R}\,\tilde f_{-}(p_0,{\bf p};x)=
v_{-}({\bf p})\,K_{i\nu\,+\frac12}(\kappa x)
+ i\kappa\,\Upsilon_{+}^{R}\,K_{i\nu\,-\frac12}(\kappa x)\end{array}\right.
\eeq
Thus $\forall\,p_0\in\mathbb R$ we find four
bispinor solutions describing the spin $\frac12$ quantum states of 
definite transverse momentum, which propagate in the Rindler wedges.
Of course they can be always set into 1:1 correspondence
with the left and right massive Weyl spinor with positive and negative
helicities which are experienced by an inertial observer, as it does.
It can be readily checked by direct inspection that the spin states do satisfy
the following orthonormality relations, viz.,
\beq
\bar u_{\pm}({\bf p})\,\gamma^0_M\,u_{\pm}({\bf p})
=\bar v_{\pm}({\bf p})\,\gamma^0_M\,v_{\pm}({\bf p})
=\kappa^2\\
\bar u_{\pm}({\bf p})\,\gamma^0_M\,\Upsilon_{\mp}^{L}=
\bar v_{\pm}({\bf p})\,\gamma^0_M\,\Upsilon_{\mp}^{R}
=0\\
\bar u_{\x\pm}({\bf p})\,\gamma^0_M\,v_{\pm}({\bf p})=
\bar v_{\pm}({\bf p})\,\gamma^0_M\,u_{\x\pm}({\bf p})
=0\\
\bar u_{\x\pm}({\bf p})\,\gamma^0_M\,\Upsilon_{\mp}^{R}=
\bar v_{\pm}({\bf p})\,\gamma^0_M\,\Upsilon_{\mp}^{L}
=0
\eeq

Thus we can build up the two sets of complex plane wave solutions of the
massive spinor covariant equation (\ref{covdiracequ}) in the right Rindler wedge
$\mathfrak W_R$ which read
\beq
\psi_{\rm p}^{\x\imath}(t,{\bf x}) &=&
\sqrt N\,\theta(x)\,\exp\{-i\nu\x t\x({\rm a}/c)+i\x{\bf p}\,\cdot\,{\bf x}_\perp\}\no
\fl
&\times&
\left\lbrace\begin{array}{cc}
u_{\x+}({\bf p})\,K_{i\nu\,-\frac12}(\kappa x)
+i\kappa\,\Upsilon_{-}^{L}\,K_{i\nu\,+\frac12}(\kappa x) & \rm for\ \imath=1\\
v_{+}({\bf p})\,K_{i\nu\,-\frac12}(\kappa x)
- i\kappa\,\Upsilon_{-}^{R}\,K_{i\nu\,+\frac12}(\kappa x) & \rm for\ \imath=2
\end{array}\right.
\eeq
\beq
\fl
\phi_{\rm p}^{\x\imath}(t,{\bf x}) &=&
\sqrt N\,\theta(x)\,\exp\{-i\nu\x t\x({\rm a}/c)+i\x{\bf p}\,\cdot\,{\bf x}_\perp\}\no
\fl
&\times&\left\lbrace\begin{array}{cc}
u_{\x-}({\bf p})\,K_{i\nu\,+\frac12}(\kappa x)
- i\kappa\,\Upsilon_{+}^{L}\,K_{i\nu\,-\frac12}(\kappa x) & \rm for\ \imath=1\\
v_{-}({\bf p})\,K_{i\nu\,+\frac12}(\kappa x)
+ i\kappa\,\Upsilon_{+}^{R}\,K_{i\nu\,-\frac12}(\kappa x) & \rm for\ \imath=2
\end{array}\right.
\eeq
$N$ being a normalization constant to be determined here below.
As a matter of fact,
the invariant inner product between any two complex solutions of the covariant Dirac equation
is defined by
\beq
(\,\psi_{\x2}\,,\,\psi_1\,)\equiv\oint_\Sigma\,V_{\alpha\mu}(x)\,\bar\psi_{\x2}(\mathrm x)\,
\gamma^{\,\alpha}\,\psi_1(\mathrm x)\,\rmd\Sigma^{\,\mu}
\eeq
where $\Sigma$ is a three dimensional future oriented space-like hyper-surface.
Again, for the initial time three dimensional hyper-surface 
in the right Rindler wedge
$\mathfrak W_R$ we get
\[
\rmd\Sigma^{0}\,=\,\frac{c^{\x2}}{{\rm a}\,x}\;\theta(x)\,{\rmd x}\,\rmd^2{\bf x}_\perp
\qquad\rmd\Sigma^\imath=0\quad(\,\imath=1,2,3\,)
\]
\[
V_{00}\,=\,\sqrt{\x-\,g}\qquad\quad
\bar \psi(x)\;=\;\psi^{\dagger}(x)\,\beta_M
\]
and consequently
\beq
(\,\psi_{\x2}\,,\,\psi_1\,)=\int_0^\infty{\rmd x}\int\rmd^2{\bf x}_\perp\;
\psi_{\x2}^{\x\dagger}(t,x,{\bf x}_\perp)
\,\psi_1(t,x,{\bf x}_\perp)
\eeq
For ${\rm p}=(\,\mu,{\bf p}\,)$ and ${\rm q}=(\,\nu,{\bf q}\,)$ we find for example
\beq
\fl\qquad
\left(\,\phi^{\x\jmath}_{\rm q}\,,\,\phi^{\x\imath}_{\rm p}\,\right)
&=& {N}\,(\x 2\pi\kappa\x)^2\;\delta_{\imath\jmath}\,\delta(\x{\bf p}-{\bf q}\x)
\,\e^{\x-\,it(\x\mu-\nu){\rm a}/c}\no
\fl\qquad
&\times& \lim_{\lambda\,\rightarrow\,0^{-}}\int_{0}^\infty{\rmd x}\,x^{-\lambda}\left\lbrace
K^\ast_{i\nu\,+\frac12}(\kappa x)\,K_{i\mu\,+\frac12}(\kappa x)\ +\
\mu\,\leftrightarrow\,\nu\right\rbrace
\label{invinnpro}
\eeq
This means that one has to first consider the integral
\cite{gradshteyn} 
\beq
\int_{0}^\infty\rmd x\;x^{\x-\x\lambda}\,
K_{i\mu\,+\frac12}(\kappa x)\,K_{i\nu\,-\frac12}(\kappa x)
=\frac{2^{\x-2-\lambda}\,\kappa^{\x\lambda-1}}{\Gamma(1-\lambda)}\no
\times\;\Gamma\left({\frac{1-\lambda+i\mu+i\nu}{2}}\right)\,
\Gamma\left({\frac{1-\lambda-i\mu-i\nu}{2}}\right)\no
\times\;\Gamma\left({\frac{2-\lambda+i\mu-i\nu}{2}}\right)\,
\Gamma\left({\frac{-\lambda-i\mu+i\nu}{2}}\right)
\qquad\quad(\,\Re\rm e\,\lambda<0\,)
\label{integral}
\eeq
and realize that the right hand side of the above equality is analytic
for $\lambda\in\mathbb R\,.$
Thus the most convenient and simple way one can understand the value of the integral appearing in the
right hand side of (\ref{invinnpro}) is in terms of  analytic regularization.
Namely, one can take the limit for $\lambda\to0^-$ of the analytic function in the right hand side 
of (\ref{integral}) that yields
\[
\fl\qquad
\int_{0}^\infty{\rmd x}\,
K_{i\mu\,+\frac12}(\kappa x)\,K_{i\nu\,-\frac12}(\kappa x)
={i\x\pi^{\x2}}/{4\kappa\cosh\ts\frac12(\mu+\nu)\sinh\frac12(\,\mu-\nu\,)}
\]
which is purely imaginary or, equivalently, manifestly antisymmetric 
under the exchange $\mu\leftrightarrow\nu\,.$
Hence, adding the complex conjugate, we immediately obtain the orthogonality
relation, viz.,
\beq
\int_{0}^\infty{\rmd x}\,
K_{i\mu\,+\frac12}(\kappa x)\,K_{i\nu\,-\frac12}(\kappa x)\;+\;{\rm c.c.}=0
\qquad\quad(\,\mu\not=\nu\,)
\eeq
Moreover, if we set $\mu-\nu=\xi+i\ve$ then we can write for $\ve\to0^+$
\beq
\fl\qquad
\int_0^\infty{\rmd x}\;K_{i\mu\,+\frac12}(\kappa x)\,K_{i\nu\,-\frac12}(\kappa x)\;
\buildrel\ve\downarrow0\over\sim\,
\frac{i\x\pi}{2\kappa\cosh\pi\mu}\cdot\frac{1}{\xi+i\ve}\no
\fl\qquad=\left\lbrace{\rm CPV}\left(\frac{1}{\xi}\right)-i\pi\delta(\xi)\right\rbrace
\frac{i\x\pi}{2\kappa\cosh\pi\mu}\,
\buildrel \ve\downarrow0\over\longrightarrow\,
\frac{\pi^2}{2\kappa}\;\delta(\x\mu-\nu\x)\;{\rm sech\pi\mu}
\eeq
in such a manner that we can eventually write
\beq
\left(\,\phi^{\x\jmath}_{\rm q}\,,\,\phi^{\x\imath}_{\rm p}\,\right)
&=&{N}\,(\x 2\pi\kappa\x)^2\;\delta_{\x\imath\jmath}\;
\delta(\x{\bf p}-{\bf p^{\,\prime}}\x)\,\delta(\,\mu-\nu\,)\;
\frac{\pi^2}{\kappa\cosh\pi\mu}\no
&=&\frac{4\pi^{\x4}N\kappa\x\hbar{\rm a}}{c\x\cosh(\pi cE/\hbar\rm a)}\;
\delta(\x{\bf p}-{\bf p^{\,\prime}}\x)\,\delta(E-E^{\,\prime}\x)\;\delta_{\x\imath\jmath}
\eeq
where $\imath,\jmath=1,2\,,$  whereas use has been made of the orthonormality relations of the spin states.
Hence we eventually come to the two complete and orthonormal sets of
normal modes for the massive spin $\frac12$ field in the right Rindler wedge
$\mathfrak W_R$
\beq
\fl\qquad
\phi^{\x\imath}_{\rm p}(t,x,{\bf x}_\perp) &=&
c\,\sqrt{\frac{c^{\x2}}{\kappa\rm a}\,\cosh\pi\mu}\;
\exp\{-i\mu\x t\x({\rm a}/c)+i\x{\bf p}\,\cdot\,{\bf x}_\perp\}\,
\varphi^{\x\imath}_{\rm p}(x)\no
\fl\qquad
&=& \sqrt{\frac{c^{\x2}}{\kappa\rm a}\,\cosh\frac{\pi cE}{\hbar\rm a}}\;
\,\exp\left\{-\,\frac{i}{\hbar}\,Et+i{\bf p}\,\cdot\,{\bf x}_\perp\right\}\,
\varphi^{\x\imath}_{\rm p}(x)
\label{normal_modes_phi}
\eeq
\beq
\fl\qquad
\varphi^{\x\imath}_{\rm p}(x)
\equiv\;\frac{\theta(x)}{2\pi^2}\;\times\;\left\lbrace\begin{array}{cc}
u_{\x-}({\bf p})\,K_{i\mu\,+\frac12}(\kappa x)
- i\kappa\,\Upsilon_{+}^{L}\,K_{i\mu\,-\frac12}(\kappa x) & \rm for\ \imath=1\\
v_{-}({\bf p})\,K_{i\mu\,+\frac12}(\kappa x)
+ i\kappa\,\Upsilon_{+}^{R}\,K_{i\mu\,-\frac12}(\kappa x) & \rm for\ \imath=2
\end{array}\right.
\eeq
\[
{\rm p}\,\equiv\,(\,\mu,{\bf p}\,)=\left(\frac{E c}{\hbar\rm a},{\bf p}\right)
\qquad\quad
{\rm q}\,\equiv\,(\,\nu,{\bf q}\,)=\left(\frac{E^{\,\prime}c}{\hbar\rm a},{\bf q}\right)
\]
\beq
\left(\,\phi^{\x\jmath}_{\rm q}\,,\,\phi^{\x\imath}_{\rm p}\,\right)\;
=\;\hbar c\,\delta(\x{\bf p}-{\bf q}\x)\,\delta(E-E^{\,\prime}\x)\;\delta_{\x\imath\jmath}
\eeq
In the very same way we get the other complete orthonormal set of bispinors
in the right Rindler wedge
$\mathfrak W_R$
\beq
\fl\qquad
\psi^{\x\imath}_{\rm p}(t,x,{\bf x}_\perp) &=&
\sqrt{\frac{c^{\x2}}{\kappa\rm a}\,\cosh\pi\mu}\;
\exp\{-i\mu\x t\x({\rm a}/c)+i\x{\bf p}\,\cdot\,{\bf x}_\perp\}\,
\chi^{\x\imath}_{\rm p}(x)\no
\fl\qquad
&=&\sqrt{\frac{c^{\x2}}{\kappa\rm a}\,\cosh\frac{\pi cE}{\hbar\rm a}}\;
\,\exp\left\{-\,\frac{i}{\hbar}\,Et+i{\bf p}\,\cdot\,{\bf x}_\perp\right\}\,
\chi^{\x\imath}_{\rm p}(x)
\label{normal_modes_psi}
\eeq
\beq
\fl\qquad
\chi^{\x\imath}_{\rm p}(x)
\equiv\;\frac{\theta(x)}{2\pi^2}\;\times\;
\left\lbrace\begin{array}{cc}
u_{\x+}({\bf p})\,K_{i\nu\,-\frac12}(\kappa x)
+i\kappa\,\Upsilon_{-}^{L}\,K_{i\nu\,+\frac12}(\kappa x) & \rm for\ \imath=1\\
v_{+}({\bf p})\,K_{i\nu\,-\frac12}(\kappa x)
- i\kappa\,\Upsilon_{-}^{R}\,K_{i\nu\,+\frac12}(\kappa x) & \rm for\ \imath=2
\end{array}\right.
\eeq
\[
{\rm p}\,\equiv\,(\,\mu,{\bf p}\,)=\left(\frac{E c}{\hbar\rm a},{\bf p}\right)
\qquad\quad
{\rm q}\,\equiv\,(\,\nu,{\bf q}\,)=\left(\frac{E^{\,\prime}c}{\hbar\rm a},{\bf q}\right)
\]
\[
\left(\,\psi^{\x\jmath}_{\rm q}\,,\,\psi^{\x\imath}_{\rm p}\,\right)=
\hbar c\,\delta(\x{\bf p}-{\bf q}\x)\,\delta(E-E^{\,\prime}\x)\;\delta_{\x\imath\jmath}
\]
The above normalized bispinors belonging to both complete and orthonormal sets 
of solutions of the covariant massive Majorana spinor wave equation are dimensionless
and turn out to be, as already noticed, equivalent and independent because
\beq
\fl\qquad
\left(\,\psi^{\x\jmath}_{\rm q}\,,\,\phi^{\x\imath}_{\rm p}\,\right)\;=\;0
\qquad\quad(\,\forall\,\mu,\nu\in\mathbb R\,,\,{\bf p},{\bf p}^{\,\prime}\in{\mathbb R}^2\,,\ \imath,\jmath=1,2\,)
\eeq
It is also important to gather that the normal modes (\ref{normal_modes_phi}) and (\ref{normal_modes_psi})
of the accelerated massive spin $\frac12$ field
do exhibit two helicity states, i.e. $\imath=1,2$.
\subsection{Canonical quantization in the Rindler wedge}
To proceed futher on, it is convenient to simplify a little bit the notations by introducing a
multi-index which collectively labels all the quantum numbers of the covariant
massive Majorana spinor solutions. To this purpose, we shall use the
indices $\iota,\varkappa,\ell,\varpi,\varrho,\varsigma,\vartheta,\ldots$
to label the quartets of quantum numbers 
\[
\mathfrak Q\;=\;\{(E,{\bf p},r)\,|\,\ E\in\mathbb R\,,\ (p_y,p_z)\in{\mathbb R}^2\,,\ \imath=1,2\}
\]
together with
\beq
\sum_{\iota\,\in\,\mathfrak Q}\;\equiv\;
\int_{-\infty}^\infty\frac{\rmd E}{\hbar c}\int\rmd^2{\bf p}\sum_{\jmath\,=\,1,2}
=\frac{\mathrm a}{c^{\x2}}\int_{-\infty}^\infty{\rmd \nu}\int\rmd^2{\bf p}\sum_{\jmath\,=\,1,2}\\
\fl\qquad
\delta_{\,\iota\varkappa}=
\hbar c\,\delta(E-E^{\,\prime}\x)\,\delta(\x{\bf p}-{\bf p^{\,\prime}}\x)\,\delta_{\x\imath\jmath}
= \frac{c^{\x2}}{\mathrm a}\,\delta(\nu-\nu^{\,\prime}\x)\,\delta(\x{\bf p}-{\bf p^{\,\prime}}\x)\,\delta^{\x\imath\jmath}=\delta^{\,\iota\varkappa}
\eeq
in such a manner
that we can write the normal modes expansion of the quantized massive Majorana
spinor hermitean fields in the right Rindler wedge
$\mathfrak W_R$ in the form
\beq
\fl
\psi(\mathrm x)=
\sum_{\iota\,\in\,\mathfrak Q}
\Big[\,a_{\x\iota}\,\psi_{\iota}(\mathrm x)
+ a_{\x\iota}^{\x\dagger}\,\psi_{\iota}^{\x\ast}(\mathrm x)\,\Big]
= \psi^{\dagger}(\mathrm x)
\qquad\quad[\,\psi\,]={\rm eV}^\frac32
\eeq
with
\[
\psi_{\iota}(\mathrm x)\,\equiv\,\psi^{\x\imath}_{\rm p}(t,x,{\bf x}_\perp)
\qquad\quad
\psi_{\iota}^{\x\ast}(\mathrm x)\,\equiv\,\bar\psi^{\x\imath}_{\rm p}(t,x,{\bf x}_\perp)\,\beta_M
\]
where the canonical anticommutation relations hold true, viz.,
\beq
\left\{a_{\varkappa}\,,\,
a_{\ell}^{\x\dagger}\right\}
=\delta^{\,\varkappa\ell}
\qquad\quad[\,a_{\varkappa}\,]={\rm cm}^\frac32
\eeq
all other anticommutators being null. Notice that the canonical anticommutation relations
drive towards the closure or completeness relation for the normal modes that reads
\begin{eqnarray}
&&\{\psi(\mathrm x)\,,\,\overline{\psi}(\mathrm y)\}\nonumber\\
&=&\sum_{\iota\,\in\,\mathfrak Q}\sum_{\varkappa\,\in\,\mathfrak Q}
\{a_{\x\iota}\,\psi_{\iota}(\mathrm x)
+ a_{\x\iota}^{\x\dagger}\,\psi_{\iota}^{\x\ast}(\mathrm x)\,,\,
a_{\x\varkappa}^{\x\dagger}\,\overline{\psi}_{\varkappa}(\mathrm y)
+ a_{\x\varkappa}\,\overline{\psi}_{\varkappa}^{\x\ast}(\mathrm y)\}\nonumber\\
&=&\sum_{\iota\,\in\,\mathfrak Q}[\,\psi_{\iota}(\mathrm x)\x\overline{\psi}_{\iota}(\mathrm y)
+ {\psi}_{\iota}^{\x\ast}(\mathrm x)\x\overline{\psi}_{\iota}^{\x\ast}(\mathrm y)\,]
= S(\mathrm x,\mathrm y)
\end{eqnarray} 
In a similar way we find
\beq
\fl
\phi(\mathrm x)=
\sum_{\iota\,\in\,\mathfrak Q}
\Big[\,b_{\x\iota}\,\phi_{\iota}(\mathrm x)
+ b_{\x\iota}^{\x\dagger}\,\phi_{\iota}^{\x\ast}(\mathrm x)\,\Big]
= \phi^{\dagger}(\mathrm x)
\qquad\quad[\,\phi\,]={\rm eV}^\frac32
\eeq
with
\[
\phi_{\iota}(\mathrm x)\,\equiv\,\phi^{\x\imath}_{\rm p}(t,x,{\bf x}_\perp)
\qquad\quad
\phi_{\iota}^{\x\ast}(\mathrm x)\,\equiv\,\overline{\phi}^{\,\imath}_{\x\rm p}(t,x,{\bf x}_\perp)\,\beta_M
\]
where the canonical anticommutation relations hold true, viz.,
\beq
\left\{b_{\varkappa}\,,\,
b_{\ell}^{\x\dagger}\right\}
=\delta^{\,\varkappa\ell}
\qquad\quad[\,a_{\varkappa}\,]={\rm cm}^\frac32
\eeq
all other anticommutators being equal to zero. 

To better understand the physical meaning of the above solutions
of the massive Majorana spinor wave equation in the Rindler
accelerated coordinates system, it is utmost convenient to
select the special class of noninertial reference frames in which $p_y=p_z=0\,.$
For any such an observer, that will be named transverse momentum rest frame
instantaneous observer, we have for example
\beq
\psi^{\x\imath}_{E}(t,x) &=&
\sqrt{\frac{c^{\x2}}{\kappa\rm a}\,\cosh\frac{\pi cE}{\hbar\rm a}}\;
\,\exp\left\{-\,\frac{i}{\hbar}\,Et\right\}\,
\chi^{\x\imath}_{\x\nu}(x)
\eeq
\beq
\fl\qquad
\chi^{\x\imath}_{\x\nu}(x)\,
=\;\frac{m}{2\pi^2}\;\theta(x)
\left\lbrace\begin{array}{cc}
\Upsilon_{+}^{L}\,K_{i\nu\,-\frac12}(\kappa x)
+i\,\Upsilon_{-}^{L}\,K_{i\nu\,+\frac12}(\kappa x) & \rm for\ \imath=L\\
\Upsilon_{+}^{R}\,K_{i\nu\,-\frac12}(\kappa x)
- i\,\Upsilon_{-}^{R}\,K_{i\nu\,+\frac12}(\kappa x) & \rm for\ \imath=R
\end{array}\right.
\eeq
\beq
cH_M&=&{\rm a}\x x\left(\alpha_M^1\,\hat P_x+mc\beta_M\right)\no
\hat P_x &=&\left\lbrace
\ba{c}
-\,i\delta_{\,+}\,-\,E/{\rm a}\x x\\
-\,i\delta_{\,-}\,+\,E/{\rm a}\x x
\ea\right.
\eeq
\[
\beta_M\,\Upsilon_{\pm}^{\,L}\;=\;-\,\Upsilon_{\mp}^{\,L}
\qquad\qquad
\beta_M\,\Upsilon_{\pm}^{\,R}\;=\;\Upsilon_{\mp}^{\,R}
\]
and consequently
\beq
H_M\,\psi^{\x\imath}_{E}(t,x)
\;=\;E\,\psi^{\x\imath}_{E}(t,x)
\qquad (\,\imath\,=\,L,R\,)\\
\Sigma_M^1\,\psi^{\x\imath}_{E}(t,x)\;=\;
\left\lbrace
\ba{cc}
+\,\psi^{\x\imath}_{E}(t,x) & {\rm for}\quad\ \imath\,=\,L\\
-\,\psi^{\x\imath}_{E}(t,x) & {\rm for}\quad\ \imath\,=\,R
\ea
\right.
\eeq
It means that, for example, the 1-particle states 
\beq
a_{\x\ell}^{\x\dagger}\,|\,0\,\rangle\;=\;|\,\ell\,+\,\rangle\;=\;|\,E\,{\bf p}\,\jmath\,\rangle
\qquad\quad(\,\forall\,\ell=E,{\bf p},\jmath\,)
\eeq
will actually describe massive, neutral, spin $\frac12$ quanta, called pseudo-particles,
with {indefinite energy}, i.e., $-\infty<E<\infty$, transverse wave numbers $p_y,p_z\in\mathbb R$ and 
definite helicities
$\jmath=1,2\,,$
i.e. spin projection along  and versus the direction of the accelerated observer,
while
\beq
b_{\x\varrho}^{\x\dagger}\,|\,0\,\rangle\;=\;|\,\varrho\,-\,\rangle\;=\;|\,E\,{\bf p}\,\imath\,\rangle
\qquad\quad(\,\forall\,\varrho=E,{\bf p},\imath\,)
\eeq
do describe massive neutral spin $\frac12$ particles with opposite helicities.
\section{The Bogolyubov Coefficients For Majorana Spinors}\label{sec:Bogolyubov}
Now we are ready to generalize the method developed\cite{LS} for the spinless and chargeless
quantum field to the neutral Majorana bispinor quantum field. To this purpose, let us
recall the normal modes expansion of a quantized Majorana Hermitean bispinor
in an inertial reference frame of the Minkowski space, i.e.,
\beq
\Psi_M\x(\mathrm X) = \sum_{\,{\bf P}\,,\,r}\,\Big[\,
A_{\,{\bf P},\,r}\,\Psi_{\,{\bf P}}^{\,r}(\mathrm X)
+ A_{\,{\bf P},\,r}^{\x\dagger}\,\Psi_{\,{\bf P}}^{\,r\x\ast}(\mathrm X)\,\Big]
= \Psi_M^{\x\dagger}\x(\mathrm X)
\eeq
\beq
\overline{\Psi}_M\x(\mathrm X) = \sum_{\,{\bf P}\,,\,r}\,\Big[\,
A^{\x\dagger}_{\,{\bf P},\,r}\,\overline{\Psi}_{\,{\bf P}}^{\,r}(\mathrm X)
+ A_{\,{\bf P},\,r}\,\overline{\Psi}_{\,{\bf P}}^{\,r\x\ast}(\mathrm X)\,\Big]
= \overline{\Psi}_M^{\,\dagger}\x(\mathrm X)
\eeq
Now we have to express the above hermitean quantum field in Rindler's curvilinear coordinates
according to (\ref{spinortransform}),
in which, however, we shall use the slightly modified and more convenient notation 
\beq
\fl
\Psi_{\,{\bf K}}^{\,r}[\,\mathrm X(\mathrm x)\,]&\equiv&
[\,(2\pi)^3\x2\x\omega_{\,\bf K}\,]^{\x-\x1/2}\;
u_{\,\mathbf K}^{\,r}(t)\,
\exp\{\,\mathrm i\x\Bbbk_t\x x+\mathrm i\x K_{\bot}\cdot x_{\bot}\}\\
\fl
\overline{\Psi}_{\,{\bf K}}^{\,r}[\,\mathrm X(\mathrm x)\,]&\equiv&
[\,(2\pi)^3\x2\x\omega_{\,\bf K}\,]^{\x-\x1/2}\;
\bar u_{\,\mathbf K}^{\,r}(t)\,
\exp\{-\,\mathrm i\x\Bbbk_t\x x - \mathrm i\x K_{\bot}\cdot x_{\bot}\}
\eeq
where
\[
\sum_{\,{\bf K}\,,\,r}\equiv\int\rmd^2{ K}_{\bot}\int_{-\infty}^{\infty}\mathrm dK
\sum_{r\,=\,\uparrow\downarrow}
\]
\[
\Bbbk_t=K\cosh\eta - {\omega_{\,\bf K}}\sinh\eta
\]
the spin states being now time dependent and denoted by
\[
 u_{\,\mathbf K}^{\,r}(t)\equiv\,D[\,\Lambda(t)\,]\,u_{\x r}(\mathrm K)
\qquad\quad \left[ \,\mathrm K^{\mu}=(\omega_{\,\bf K},\mathbf K)\,,\,r=\uparrow\downarrow\,\right] 
\]
where $u_{\x r}(\x \mathrm K\x)\ (r=\uparrow,\downarrow)$ are the usual spin states, defined as%
\begin{equation}
\left.
\ba{c}
u_{\x r}(\x \mathrm K\x)\equiv 2m\x(2\omega_{\,\bf K}+2m)^{-\frac12}\,
{\mathcal E}_M^{\x+}\x(\x \mathrm K \x)\,\xi_{\,r}\\
u^{\ast}_{\x r}(\x \mathrm K\x)\equiv 2m\,(2\omega_{\,\bf K}+2m)^{-\frac12}\,
{\mathcal E}_M^{\x-}\x(\x \mathrm K \x)\,\xi^{\ast}_{\,r}
\ea\;\right\rbrace
\qquad(\,r\,=\,\uparrow\,,\,\downarrow)
\label{spinstates} 
\end{equation}
which are moreover the eigenstates of the positive energy projector
\beq
&& {\mathcal E}_M^{\x\pm}\x(\x {\mathrm K} \x)=\left(\x m\pm\gamma^{\,\mu}_M\,{\mathrm K}_\mu\right)\x/\x{2m}
\qquad\qquad\,\left(\,\mathrm K_0=\omega_{\,\bf K}\,\right)\\
&& {\mathcal E}_M^{\x+}\x(\x \mathrm K\x)\,u_{\x r}\x(\x \mathrm K \x) = u_{\x r}\x(\x \mathrm K\x)
\qquad\qquad\ \ \quad\quad(\,r\,=\,\uparrow\,,\,\downarrow)
\eeq
where 
\beq
\xi_{\x\uparrow}\equiv\left\lgroup
\ba{c}1\\-i\\0\\0\ea\right\rgroup
\qquad\quad
\xi_{\x\downarrow}\equiv\left\lgroup
\ba{c}0\\0\\1\\i\ea\right\rgroup
\eeq
are the eigen-bispinors of the rest frame Majorana Hamiltonian, both with eigenvalue $m$.

The canonical anticommutation relations hold true
 \beq
\{A_{\,{\bf P},\,r}\,,\,A_{\,{\bf K},\,s}\}=
\left\lbrace A^{\x\dagger}_{\,{\bf P},\,r}\,,\,A^{\x\dagger}_{\,{\bf K},\,s}\right\rbrace =0
\eeq
\beq
\left\lbrace A_{\,{\bf P},\,r}\,,\,A^{\x\dagger}_{\,{\bf K},\,s}\right\rbrace \ =\
\delta_{\,rs}\;\delta\left({\bf P}-{\bf K}\right)
\no
{\bf P}\x,\x{\bf K}\in{\mathbb R}^3\,,\ r\x,\x s=\uparrow\downarrow
\eeq
Notice that from the general invariant and time-independent normalization in the space-like region of
the right Rindler's wedge
\beq
\left(\,\Psi_{\,{\bf P}}^{\,s}\,,\,\Psi_{\,{\bf K}}^{\,r}\,\right)
&=&\int\mathrm d^3{X}\ 
\overline{\Psi}_{\,{\bf P}}^{\,s}(\tau,X)\,\beta_M\,
\Psi_{\,{\bf K}}^{\,r}(\tau,X)\nonumber\\
&=&\int_{0}^{\infty}\mathrm dx\int\mathrm d^{2}x_{\bot}\,
\overline{\Psi}_{\,{\bf P}}^{\,s}[\,\mathrm X(\mathrm x)\,]\,\beta_M\,
\Psi_{\,{\bf K}}^{\,r}[\,\mathrm X(\mathrm x)\,]\nonumber\\
&=&\delta^{\,rs}\;\delta({\bf P}-{\bf K})
\eeq
we can readily obtain the normal modes expansions of the observables
involving the Majorana massive spinor field.
The Majorana field equation,  which is satisfied by the real self-conjugated 
bispinor field for any inertial observer, reads
\begin{equation}
\gamma_M^{\x\alpha}\x\pa_\alpha\Psi_M(\mathrm X) + im\,\Psi^{\,c}_M(\mathrm X)=0
\qquad\quad
\Psi_M=\Psi^{\,c}_M=\Psi^{\ast}_{M}
\end{equation}
To proceed further on, consider the basic integrals
\begin{eqnarray*}
\fl
&& I_{i\nu\pm\frac12}(\kappa,P)\,\equiv\,\int_{0}^{\infty}\mathrm dx\;\mathrm e^{-\,iPx}\,K_{i\nu\pm\frac12}(\kappa x)
\,=\,\frac{\pi\rm cosec(i\pi\nu\pm\pi/2)}{\surd(-\kappa^2-P^2)}\\
\fl&\times&\left[ \,\kappa^{-i\nu\mp\frac12}\left( 
\sqrt{-\kappa^2-P^2} + iP\right)^{i\nu\pm\frac12}
 - \kappa^{i\nu\pm\frac12}\left( \sqrt{-\kappa^2-P^2} + iP\right)^{-i\nu\mp\frac12}\,\right]\\
\fl&=&\frac{\mp\x\pi i}{\omega_{\,\bf P}\x\cosh(\pi\nu)}\,
\left[ \,\mathrm e^{-\,\frac12\pi\nu\pm\,\frac14\pi i}\,\left(\frac{P + \omega_{\,\bf P}}{\kappa}
\right)^{i\nu\pm\frac12}
-\;\mathrm e^{\,\frac12\pi\nu\mp\,\frac14\pi i}\,\left(\frac{P + \omega_{\,\bf P}}{\kappa}
\right)^{-\,i\nu\mp\frac12}\,\right]
\end{eqnarray*}
so that if we set
\begin{eqnarray}
 \omega_{\,\bf P}=\kappa\cosh\theta\qquad\quad
P=\kappa\sinh\theta\label{theta1} \\
\kappa^{2}=\omega_{\,\bf P}^{2} - P^{2}\qquad\quad
\theta={\rm Arth}(\x P/\omega_{\,\bf P}\x)\label{theta2} 
\end{eqnarray} 
we can recast the basic integral in the form \cite{gradshteyn}
 \begin{eqnarray}
I_{i\nu\pm\frac12}(\kappa,P)&=&
\frac{\pi}{2i\omega_{\,\bf P}\x\cosh(\pi\nu)}\,
\left[\,\left( i\x\mathrm e^{\x\theta}\x\right) ^{i\nu\pm1/2}
- \left( i\x\mathrm e^{\x\theta}\x\right) ^{-\x i\nu\mp1/2}\,\right]\no
&=& \frac{\pi}{\kappa\cosh\theta\x\cosh(\pi\nu)}\,\sin\left[ \,
\theta\nu+\frac{\pi i\nu}{2}\pm\frac{i\theta}{2}\pm \frac{\pi}{4}\,\right] 
 \end{eqnarray} 
The accelerated noninertial observer does instead employ e.g. the other complete and orthonormal set 
(\ref{normal_modes_psi}) of bispinors
in the right Rindler wedge
$\mathfrak W_R$
\begin{eqnarray*}
\fl
\psi^{\x\imath}_{\mathrm p}(\mathrm x) &=&
\sqrt{\frac{1}{\kappa\rm a}\,\cosh\pi\mu}\;
\exp\{-i{\rm a}\x\mu\x t+i\x{\bf p}\,\cdot\,{\bf x}_\perp\}\,
\chi^{\x\imath}_{\rm p}(x)\\
\fl
\chi^{\x\imath}_{\rm p}(x)
&\equiv&\frac{1}{2\pi^2}\;\times\;
\left\lbrace\begin{array}{cc}
u_{\x+}({\bf p})\,K_{i\mu\,-\frac12}(\kappa x)
+i\kappa\,\Upsilon_{-}^{L}\,K_{i\mu\,+\frac12}(\kappa x) & \rm for\ \imath=1\\
v_{+}({\bf p})\,K_{i\mu\,-\frac12}(\kappa x)
- i\kappa\,\Upsilon_{-}^{R}\,K_{i\mu\,+\frac12}(\kappa x) & \rm for\ \imath=2
\end{array}\right.\quad (x\ge0)
\end{eqnarray*}
\[
{\rm p}\,\equiv\,(\,\mu,{\bf p}\,)=\left(\frac{E}{\rm a},{\bf p}\right)
\qquad\quad
{\rm q}\,\equiv\,(\,\nu,{\bf q}\,)=\left(\frac{E^{\,\prime}}{\rm a},{\bf q}\right)
\]
\[
\left(\,\psi^{\x\jmath}_{\rm q}\,,\,\psi^{\x\imath}_{\rm p}\,\right)=
\frac{1}{\mathrm a}\,\delta(\x{\bf p}-{\bf q}\x)\,\delta(\mu - \nu\x)\;\delta^{\x\imath\jmath}
\]
The Bogolyubov coefficients for the Hermitean self-conjugated Majorana quantized
bispinor field are thereby defined as follows: namely,
\begin{eqnarray}
\alpha(\mathbf P,r\x;\mathrm p,\imath)&\equiv&
\left(\,\Psi_{\x{\bf P}}^{\x r}\,,\,\psi^{\x\imath}_{\x\mathrm p}\,\right) =
2\pi\,\delta(\mathbf P_\bot -\mathbf p)\nonumber\\
&\times&\sqrt{\frac{\cosh\pi\mu}{4\pi\kappa\mathrm a\x\omega_{\,\bf P}}}\;
\int_{0}^{\infty}
\bar u^{\x r}_{\x\mathbf P}(0)\beta_{M}\chi^{\x\imath}_{\x\rm p}(x)\;
\mathrm e^{-iPx}\,\mathrm dx\nonumber\\
&=& \delta(\mathbf P_\bot -\mathbf p)\;\sqrt{\frac{c^{\x3}}{2\pi\mathrm a\x\omega_{\,\bf P}}}
\nonumber\\
&\times&\left\lbrace \mathbb R^{\x\imath\,r}_{\x\mathrm p\mathbf P}\,I_{i\mu-\frac12}(\kappa,P)
+\overline{\mathbb R}^{\x\imath\,r}_{\x\mathrm p\mathbf P}\,I_{i\mu+\frac12}(\kappa,P)\right\rbrace 
\end{eqnarray} 
where we have introduced the spin correlation matrices
\begin{eqnarray}
 \mathbb R^{\x\imath\,r}_{\x\mathrm p\mathbf P}=
\frac{1}{\pi}\,\left\lgroup
\begin{array}{cc}
\bar u^{\,\uparrow}_{\,\mathbf P}\beta_{M} u_{\x+}({\bf p}) & 
\bar u^{\,\downarrow}_{\,\mathbf P}\beta_{M} u_{\x+}({\bf p})\\
\bar u^{\,\uparrow}_{\,\mathbf P}\beta_{M} v_{\x+}({\bf p}) & 
\bar u^{\,\downarrow}_{\,\mathbf P}\beta_{M} v_{\x+}({\bf p})
\end{array}\right\rgroup\,\sqrt{\frac{\cosh\pi\mu}{2\kappa}}\\
\overline{\mathbb R}^{\x\imath\,r}_{\x\mathrm p\mathbf P}= 
\frac{\mathrm i}{\pi}\,\left\lgroup
\begin{array}{cc}
\bar u^{\,\uparrow}_{\,\mathbf P}\beta_{M} \Upsilon_{-}^{L} & 
\bar u^{\,\downarrow}_{\,\mathbf P}\beta_{M} \Upsilon_{-}^{L}\\
-\,\bar u^{\,\uparrow}_{\,\mathbf P}\beta_{M} \Upsilon_{-}^{R} & 
-\,\bar u^{\,\downarrow}_{\,\mathbf P}\beta_{M} \Upsilon_{-}^{R}
\end{array}\right\rgroup\,\sqrt{\textstyle\frac12\,\kappa\cosh\pi\mu}
\end{eqnarray}
Notice that we have suitably factorized the classical volume factor
\[\delta(\mathbf P_\bot -\mathbf p)\;\sqrt{\frac{c^{\x3}}{2\pi\mathrm a\x\omega_{\,\bf P}}}\]
just like in the spinless {case\cite{LS}}.
Then the Bogolyubov coefficients read
\begin{eqnarray}
&&\alpha(\mathbf P,\uparrow\x;\mathrm p,1) =
\left(\,\Psi_{\x{\bf P}}^{\x\uparrow}\,,\,\psi^{\x1}_{\x\mathrm p}\,\right)\\
&=& -\,{\textstyle\frac{1}{2}}\left(1-{\rm i}\right)\, \delta(\mathbf P_{\bot}-\mathbf p) \,
\sqrt{\frac{\cosh\pi  \nu}{(2\pi)^{3}\kappa\mathrm a \x\omega_{\,\bf P}(m+\x\omega_{\,\bf P})}} \no
&\times& \Big\{\left[\,m^2+(m+p_y)\x\omega_{\,\bf P} + p_y^2+p_x \left(m+p_y+i p_z\right)+
p_z(p_z + i \omega_{\,\bf P})\,\right]\no
&\times& {I}_{i\nu-\frac12}(\kappa,P) - i \kappa  \left(m+\omega_{\,\bf P} -p_x+p_y+i p_z\right) 
{I}_{i\nu+\frac12}(\kappa,P) \Big\}
\nonumber\\
&& \alpha(\mathbf P,\uparrow\x;\mathrm p,2) =
\left(\,\Psi_{\x{\bf P}}^{\x\uparrow}\,,\,\psi^{\x2}_{\x\mathrm p}\,\right)\\
&=& -\,{\textstyle\frac{1}{2}}\left(1-{\rm i}\right)\, \delta(\mathbf P_{\bot}-\mathbf p) \,
\sqrt{\frac{\cosh\pi  \nu}{(2\pi)^{3}\kappa\mathrm a \x\omega_{\,\bf P}(m+\x\omega_{\,\bf P})}} \no
&\times& \Big\{\left[\,m^2 + (m-p_y)\x\omega_{\,\bf P} + p_y^2 + p_x 
(m-p_y+i p_z) + p_z(p_z + i\x\omega_{\,\bf P})\,\right] \no
&\times& {I}_{i\nu-\frac12}(\kappa,P) - i \kappa  \left(m+\omega_{\,\bf P} - p_x - p_y + i p_z\right)
{I}_{i\nu+\frac12}(\kappa,P)\Big\}
\nonumber\\
&&\alpha(\mathbf P,\downarrow\x;\mathrm p,1) =
\left(\,\Psi_{\x{\bf P}}^{\x\downarrow}\,,\,\psi^{\x1}_{\x\mathrm p}\,\right)\\
&=& -\,{\textstyle\frac{1}{2}}\left(1+{\rm i}\right)\, \delta(\mathbf P_{\bot}-\mathbf p) \,
\sqrt{\frac{\cosh\pi  \nu}{(2\pi)^{3}\kappa\mathrm a \x\omega_{\,\bf P}(m+\x\omega_{\,\bf P})}}  \no
&\times& \Big\{\left[\,m^2 + (m - p_y)\x\omega_{\,\bf P} + p_y^2 +p_x\left(m-p_y-i p_z\right)
+ p_z\left(p_z-i \omega_{\,\bf P}\right)\,\right]\no
&\times& {I}_{i\nu-\frac12}(\kappa,P) 
- i \kappa  \left(m + \omega_{\,\bf P} -p_x-p_y-i p_z\right){I}_{i\nu+\frac12}(\kappa,P)\Big\}
\nonumber\\
&&\alpha(\mathbf P,\downarrow\x;\mathrm p,2) =
\left(\,\Psi_{\x{\bf P}}^{\x\downarrow}\,,\,\psi^{\x2}_{\x\mathrm p}\,\right)\\
&=& -\,{\textstyle\frac{1}{2}}\left(1+{\rm i}\right)\, \delta(\mathbf P_{\bot}-\mathbf p) \,
\sqrt{\frac{\cosh\pi  \nu}{(2\pi)^{3}\kappa\mathrm a \x\omega_{\,\bf P}(m+\x\omega_{\,\bf P})}} \no
&\times& \Big\{\left[\,m^2 + (m + p_y)\x\omega_{\,\bf P} + p_y^2 + p_x \left(m+p_y-i p_z\right)
+ p_z (p_z -i \omega_{\,\bf P})\,\right]\no
&\times& {I}_{i\nu-\frac12}(\kappa,P)
- i \kappa  \left(m+\omega_{\,\bf P} -p_x+p_y-i p_z\right){I}_{i\nu+\frac12}(\kappa,P)\Big\}\nonumber
\end{eqnarray}
Now it is possible to verify that we have the following sum rule for the dimensionless coefficients
\begin{eqnarray}
&& \sum_{\imath=1,2}\,\sum_{r=\uparrow\downarrow}\left\vert\,
\mathbb R^{\x\imath\,r}_{\x\mathrm p\mathbf P}\,I_{i\nu-\frac12}(\kappa,P)
+\overline{\mathbb R}^{\x\imath\,r}_{\x\mathrm p\mathbf P}\,I_{i\nu+\frac12}(\kappa,P)\,
\right\vert^{2}\no
&=& \frac{\kappa^{2}}{\pi^{2}}\,\cosh(\pi\nu)\,
\Big\lbrace \mathrm e^{\x-\x\theta}\,\vert\,{I}_{i\nu+\frac12}(\kappa,P)\,\vert^{2}
+ \mathrm e^{\x\theta}\,\vert\,{I}_{i\nu-\frac12}(\kappa,P)\,\vert^{2}\no
&+& \mathrm i\left[ \,I^{\x\ast}_{i\nu+\frac12}(\kappa,P)\,I_{i\nu-\frac12}(\kappa,P)
- I_{i\nu+\frac12}(\kappa,P)\,I^{\x\ast}_{i\nu-\frac12}(\kappa,P)\,\right] \Big\rbrace \no
&=& \quad\frac{2}{1+\mathrm e^{\x2\pi\nu}}\;\equiv\; \bar{N}_{\nu}
\end{eqnarray}
where the hyperbolic parameter $\theta$ -- call it {\sl rapidity} -- is defined by 
eq.s (\ref{theta1}) and  (\ref{theta2}).

In order to determine the complementary Bogolyubov coefficients $\beta(\mathbf P,r\x;\mathrm p,\imath)$,
we need to employ Minkowskian normal modes of negative frequencies that read
\begin{eqnarray}
\beta(\mathbf P,r\x;\mathrm p,\imath)&\equiv&
\left(\,\Psi_{\x{\bf P}}^{\x r\x\ast}\,,\,\psi^{\x\imath}_{\x\mathrm p}\,\right) =
2\pi\,\delta(\mathbf P_\bot +\x\mathbf p)\nonumber\\
&\times&\sqrt{\frac{\cosh\pi\mu}{4\pi\kappa\mathrm a\x\omega_{\,\bf P}}}\;
\int_{0}^{\infty}
[\,\bar u^{\x r}_{\x\mathbf P}(0)\,]^{\ast}\beta_{M}\chi^{\x\imath}_{\x\rm p}(x)\;
\mathrm e^{iPx}\,\mathrm dx\nonumber\\
&=& \delta(\mathbf P_\bot +\x\mathbf p)\;\sqrt{\frac{c^{\x3}}{2\pi\mathrm a\x\omega_{\,\bf P}}}
\nonumber\\
&\times&\left\lbrace \mathbb S^{\x\imath\,r}_{\x\mathrm p\mathbf P}\,I_{i\mu+\frac12}^{\ast}(\kappa,P)
+\overline{\mathbb S}^{\x\imath\,r}_{\x\mathrm p\mathbf P}\,I_{i\mu-\frac12}^{\ast}(\kappa,P)\right\rbrace 
\end{eqnarray} 
where we have introduced the spin correlation matrices
\begin{eqnarray}
 \mathbb S^{\x\imath\,r}_{\x\mathrm p\mathbf P}=
\frac{1}{\pi}\,\left\lgroup
\begin{array}{cc}
\bar u^{\,\uparrow\x\ast}_{\,\mathbf P}\beta_{M} u_{\x+}({\bf p}) & 
\bar u^{\,\downarrow\x\ast}_{\,\mathbf P}\beta_{M} u_{\x+}({\bf p})\\
\bar u^{\,\uparrow\x\ast}_{\,\mathbf P}\beta_{M} v_{\x+}({\bf p}) & 
\bar u^{\,\downarrow\x\ast}_{\,\mathbf P}\beta_{M} v_{\x+}({\bf p})
\end{array}\right\rgroup\,\sqrt{\frac{\cosh\pi\mu}{2\kappa}}\\
\overline{\mathbb S}^{\x\imath\,r}_{\x\mathrm p\mathbf P}= 
\frac{\mathrm i}{\pi}\,\left\lgroup
\begin{array}{cc}
\bar u^{\,\uparrow\x\ast}_{\,\mathbf P}\beta_{M} \Upsilon_{-}^{L} & 
\bar u^{\,\downarrow\x\ast}_{\,\mathbf P}\beta_{M} \Upsilon_{-}^{L}\\
-\,\bar u^{\,\uparrow\x\ast}_{\,\mathbf P}\beta_{M} \Upsilon_{-}^{R} & 
-\,\bar u^{\,\downarrow\x\ast}_{\,\mathbf P}\beta_{M} \Upsilon_{-}^{R}
\end{array}\right\rgroup\,\sqrt{\textstyle\frac12\,\kappa\cosh\pi\mu}
\end{eqnarray}
Thus we get for example
\begin{eqnarray}
&&\beta(\mathbf P,\uparrow\x;\mathrm p,1) =
\left(\,\Psi_{\x{\bf P}}^{\x\uparrow\x\ast}\,,\,\psi^{\x1}_{\x\mathrm p}\,\right)\\
&=& {\textstyle\frac{1}{2}}\left(1+{\rm i}\right)\, \delta(\mathbf P_{\bot}+\mathbf p) \,
\sqrt{\frac{\cosh\pi  \nu}{(2\pi)^{3}\kappa\mathrm a \x\omega_{\,\bf P}(m+\x\omega_{\,\bf P})}} \no
&\times& \Big\{\left[\,m^2+(m-p_y)\x\omega_{\,\bf P} + p_y^2+p_x \left(m-p_y-i p_z\right) +
p_z(p_z - i \omega_{\,\bf P})\,\right]\no
&\times& {I}_{i\nu+\frac12}^{\ast}(\kappa,P) + i \kappa  \left(m+\omega_{\,\bf P} -p_x-p_y-i p_z\right) 
{I}_{i\nu-\frac12}^{\ast}(\kappa,P) \Big\}\nonumber
\end{eqnarray}
and quite closer formulas for the three remaining components. Once again, just like in the
case of the $\alpha-$type Bogolyubov coefficients,
it is straightforward although tedious to check the following 
quite remarkable sum rule for the dimensionless coefficients
\begin{eqnarray}
&& \sum_{\imath=1,2}\,\sum_{r=\uparrow\downarrow}\left\vert\,
\mathbb S^{\x\imath\,r}_{\x\mathrm p\mathbf P}\,I_{i\nu+\frac12}^{\ast}(\kappa,P)
+\overline{\mathbb S}^{\x\imath\,r}_{\x\mathrm p\mathbf P}\,I_{i\nu-\frac12}^{\ast}(\kappa,P)\,
\right\vert^{2}\no
&=& \frac{\kappa^{2}}{\pi^{2}}\,\cosh(\pi\nu)\,
\Big\lbrace \mathrm e^{\x-\x\theta}\,\vert\,{I}_{i\nu-\frac12}(\kappa,P)\,\vert^{2}
+ \mathrm e^{\x\theta}\,\vert\,{I}_{i\nu+\frac12}(\kappa,P)\,\vert^{2}\no
&+& \mathrm i\left[ \,I^{\x\ast}_{i\nu-\frac12}(\kappa,P)\,I_{i\nu+\frac12}(\kappa,P)
- I_{i\nu-\frac12}(\kappa,P)\,I^{\x\ast}_{i\nu+\frac12}(\kappa,P)\,\right] \Big\rbrace \no
&=& \quad\frac{2}{1+\mathrm e^{\x-\x2\pi\nu}}\;\equiv\; {N}_{\nu}
\end{eqnarray}
which drives to the expected Pauli principle and spin sum relations
\begin{equation}
 {N}_{\nu}+\bar{N}_{\nu}=2\mathrm S+1
 \qquad\quad
 \forall\,\nu=\frac{Ec}{\hbar\rm a}
 \qquad\quad
 \left(\,\mathrm S=\textstyle\frac12\,\right)
\end{equation} 
and provides the crucial test for the correctness of our long and
delicate derivation.
\section{Discussion And Conclusions}
In this paper we have explicitly derived the Bogolyubov coefficients which express the probability
amplitude for particle creation out of the inertial vacuum for a neutral, massive, spin $ \frac12 $
field in a Rindler coordinate system, i.e., as experienced by some uniformly accelerated observer.
%
%
Our method for the evaluation of the probability amplitudes is nothing but a straightforward though
highly non-trivial application 
of the very definition. In particular, it relies on the explicit knowledge of the Fulling-type modes for the Majorana field. 
The very evaluation of the Bogolyubov coefficients, in the case of spinless particles, has often been carried out
in the literature by indirect methods involving, sometimes, somewhat ambiguous mathematical trickeries. 
This, in turn, sparked some controversy regarding the validity of such derivations, raised by 
\cite{fedotov} et seq. Hence, it is worthwhile to stress that our method is exempt from the criticism raised in those references, because of the direct and straightforward techniques adopted, therefore directly addressing this issue.

The Bogolyubov coefficients can be eventually expressed in closed parametric form in terms of
the rapidity $ \theta={\rm Arsh}(P/\kappa) $, where $ P $ and $ \kappa $ are the longitudinal and
transverse momenta of the particle in an inertial reference frame --
transverse and longitudinal are understood with respect the the direction of the acceleration.

Among the possible applications of our results, we briefly mention the simple situation of a de Sitter universe with constant Hubble parameter. We consider an observer located at fixed distance from (but inside) the Hubble sphere of another, geodesic, observer. The first observer is then non-geodesic, and is known as a \emph{Kodama observer} \cite{Casadio}. Then our considerations can be applied locally around the location of the non-geodesic observer: the relation between tangent spaces at that point, from the point of view of the two different frames, is precisely the same as Minkowski-Rindler. The relative acceleration would then be given by the acceleration of the Hubble horizon, namely 
\[
{\rm a}_{\x\rm cosmic}=
cH\;\approx\;2.1\times10^{\x-\x9}\ \rm m\ s^{\x-2}
\]
Then the spin $ \frac12 $ WIMP candidates would be arranged in a Fermi-Dirac
distribution at the thermal equilibrium with an absolute temperature
\[
T_{\x\rm cosmic}=\frac{\hbar H}{2\pi k_{\x\rm B}}\;\approx\;8.0\times10^{\x-\x30}\ ^\circ\rm K
\]
the Unruh cosmic temperature, in agreement with \cite{Casadio}.

On the one hand, taking into account that $ k_{\x\rm B}T_{\x\rm cosmic} \approx 1\times10^{\x-\x33}$ eV,
it appears that the energy density induced by the Unruh effect is very much suppressed. 
Instead, in order to obtain an appreciable contribution to the energy density, one would need to consider  
an acceleration much larger than $a_{\x\rm cosmic}$. It must be noted that, in the case of two relatively accelerating galaxies, these would be both geodesic observers. Nevertheless, it can be shown \cite{Strominger} that the effect still exists in that context, the temperature being the same as the one obtained above (for a mutual distance of the order of the Hubble radius).\\
On the other hand, a quite close phenomenon occurs, {\em mutatis mutandis}, in the vicinity 
of a black-hole horizon \cite{NVW}, so that the present calculation might be also useful in the
Quantum Gravity framework. 

Last but not least, our method of calculation can be further applied to the
evaluation of the Bogolyubov coefficients for the emission of charged spin $ \frac12 $ pairs and
vector particles in a uniformly accelerated reference frame, which are still unknown.
\section*{Acknowledgements}
R.S. wishes to acknowledge the support of the Istituto Nazionale di Fisica Nucleare,
Iniziativa Specifica PI14, that contributed to the successful completion of this project. The work of P.L. is supported by the DOE under grant DE-FG$02$-$96$ER$40959$

\appendix 
\section{Tetrads and the spin-connection for the Majorana field}\label{sec:tetrads}
This section provides a concise summary of our conventions for the tetrads and the spin connection of the Majorana field.
First we define
\begin{eqnarray*}
\fl\qquad
\mathrm X^{\alpha}(\mathrm x^{\x\mu}) =
\left\lbrace \begin{array}{c}
c\x\tau(ct,x,y,z)\\
X(ct,x,y,z)\\
Y(ct,x,y,z)\\
Z(ct,x,y,z)\end{array}\right.
\qquad\qquad\quad\
\left( \frac{\partial\mathrm X^{\alpha}}{\partial\mathrm x^{\x\mu}}\right) \equiv
\zeta^{\x\alpha}_{\x\mu}(\mathrm x)=\Lambda^{\alpha}_{\beta}(\mathrm x)V^{\beta}_{\mu}(\mathrm x)\\
\mathrm d s^2=\eta_{\alpha\beta}\,{\rm dX^{\alpha}dX^{\beta}}=
\eta_{\alpha\beta}\,V^{\alpha}_{\mu}(\mathrm x)V^{\beta}_{\nu}(\mathrm x)\,{\rm dx^{\mu}dx^{\nu}}
\equiv g_{\mu\nu}(\mathrm x)\,{\rm dx^{\mu}dx^{\nu}}
\end{eqnarray*}
where the local Lorentz transformation $\Lambda(\rm x)$ is a rank-four square matrix belonging to
the irreducible vector representation $D_{\frac12\frac12}$ of the Lorentz group,
its matrix elements $\Lambda^{\alpha}_{\beta}(\rm x)$ being dependent upon the curvilinear coordinates.
Next we set \cite{weinberg}
\begin{eqnarray*}
\Psi_M[\,\mathrm X(\mathrm x)\,]\equiv D[\,\Lambda(\mathrm x)\,]\,\psi_{M}(\mathrm x)\\
D[\,\Lambda(\mathrm x)\,]=\exp\left\lbrace 
- {\textstyle\frac12}\,\mathrm i\,\sigma^{\,\alpha\beta}_{M}\,\omega_{\,\alpha\beta}(\mathrm x)\right\rbrace
\qquad\quad
\sigma^{\,\alpha\beta}_{M} \equiv 
{\textstyle\frac14}\,\mathrm i\,[\,\gamma^{\,\alpha}_{M}\,,\,\gamma^{\,\beta}_{M}\,]\\
\frac{\partial}{\partial\rm X^{\alpha}}=
\left( \frac{\partial\x\rm x^{\mu}}{\partial\rm X^{\alpha}}\right) 
\frac{\partial}{\partial\x\rm x^{\mu}}\equiv
\Lambda_{\alpha}^{\beta}(\mathrm x)V^{\mu}_{\beta}[\,\mathrm X(\mathrm x)\,]\,\frac{\partial}{\partial\x\rm x^{\mu}}\\
V^{\mu}_{\alpha}[\,\mathrm X(\mathrm x)\,]\,V^{\alpha}_{\nu}(\mathrm x)=\delta^{\mu}_{\nu}
\qquad\quad
V^{\mu}_{\alpha}[\,\mathrm X(\mathrm x)\,]\,V^{\beta}_{\mu}(\mathrm x)=\delta^{\beta}_{\alpha}
\end{eqnarray*}
where $D[\,\Lambda(\mathrm x)\,]$ denotes the four dimensional spinor
local representation 
$D[\,\Lambda(\mathrm x)\,]= D_{\frac12\x0}[\,\Lambda(\mathrm x)\,]\oplus D_{0\x\frac12}[\,\Lambda(\mathrm x)\,]$ 
of the Lorentz group in the purely imaginary Majorana representation for the gamma matrices%
\beq
\gamma^{\,0}_M=
\left\lgroup\ba{cc}
-\,\sigma_{2} & 0\\
0 & \sigma_{2}\ea\right\rgroup
\qquad 
\gamma^{\,1}_M=
\left\lgroup\ba{cc}
-\,i\sigma_{3} & 0\\
0 & -\,i\sigma_{3}\ea\right\rgroup
\no
\gamma^{\,2}_M=
\left\lgroup\ba{cc}
0 & \sigma_{2}\\
-\,\sigma_{2} & 0\ea\right\rgroup
\qquad
\gamma^{\,3}_M=
\left\lgroup\ba{cc}
i\sigma_{1} & 0\\
0 & i\sigma_{1}\ea\right\rgroup\nn.
\eeq 
$D[\,\Lambda(\mathrm x)\,]$  satisfies
\[D^{-1}[\,\Lambda(\mathrm x)\,]=D[\,\Lambda^{-1}(\mathrm x)\,]=
\exp\left\lbrace 
{\textstyle\frac12}\,\mathrm i\,\sigma^{\,\alpha\beta}_{M}\,\omega_{\,\alpha\beta}(\mathrm x)\right\rbrace\]
Then we get
\begin{eqnarray*}
&& \gamma_M^\alpha\,\frac{\rm i\x\partial}{\partial\mathrm X^{\alpha}}\,\Psi_M(\mathrm X)=
\gamma_M^\alpha\,\Lambda_{\alpha}^{\beta}(\mathrm x)V^{\mu}_{\beta}[\,\mathrm X(\mathrm x)\,]\,
\frac{\rm i\x\partial}{\partial\x\rm x^{\mu}}
\left\lbrace D[\,\Lambda(\mathrm x)\,]\,\psi_{M}(\mathrm x)\right\rbrace\\
&=&\Lambda_{\alpha}^{\beta}(\mathrm x)\,V^{\mu}_{\beta}[\,\mathrm X(\mathrm x)\,]\,
\gamma_M^\alpha\left\lbrace 
{\rm i\x\partial_{\mu}}D[\,\Lambda(\mathrm x)\,]\right\rbrace
\psi_{M}(\mathrm x)\\
&+& \Lambda_{\alpha}^{\beta}(\mathrm x)\,V^{\mu}_{\beta}[\,\mathrm X(\mathrm x)\,]\,
\gamma_M^\alpha\,D[\,\Lambda(\mathrm x)\,]\,
{\rm i\x\partial_{\mu}}\psi_{M}(\mathrm x)
\end{eqnarray*}
Thus, if we multiply from the left by the inverse matrix $D^{-1}[\,\Lambda(\mathrm x)\,]$
we find
\begin{eqnarray*}
m\,\psi_{M}^{\x c}(\mathrm x)&=& \Lambda_{\alpha}^{\beta}(\mathrm x)\,V^{\mu}_{\beta}[\,\mathrm X(\mathrm x)\,]\,
D^{-1}[\,\Lambda(\mathrm x)\,]\,\gamma_M^\alpha\left\lbrace 
{\rm i\x\partial_{\mu}}D[\,\Lambda(\mathrm x)\,]\right\rbrace\psi_{M}(\mathrm x)\\
&+&\Lambda_{\alpha}^{\beta}(\mathrm x)\,V^{\mu}_{\beta}[\,\mathrm X(\mathrm x)\,]\,
D^{-1}[\,\Lambda(\mathrm x)\,]\,\gamma_M^\alpha\,D[\,\Lambda(\mathrm x)\,]\,
{\rm i\x\partial_{\mu}}\psi_{M}(\mathrm x)
\end{eqnarray*}
and from the well known transformation properties of the gamma matrices
\[D^{-1}[\,\Lambda(\mathrm x)\,]\,\gamma_M^\alpha\,D[\,\Lambda(\mathrm x)\,]=
\Lambda^{\alpha}_{\delta}(\mathrm x)\,\gamma_M^\delta
\qquad\quad
\Lambda_{\eta}^{\alpha}(\mathrm x)\,\Lambda^{\eta}_{\beta}(\mathrm x)=\delta_{\beta}^{\alpha}\]
together with the definition
\begin{equation}
\fl
\Lambda_{\alpha}^{\beta}(\mathrm x)\,V^{\mu}_{\beta}[\,\mathrm X(\mathrm x)\,]\, D^{-1}[\,\Lambda(\mathrm x)\,]\,\gamma_M^\alpha\left\lbrace {\rm i\x\partial_{\mu}}D[\,\Lambda(\mathrm x)\,]\right\rbrace
\equiv V^{\mu}_{\alpha}(\mathrm x)\,\gamma_M^\alpha\,\Gamma_{\mu}(\mathrm x)
\end{equation} 
we eventually come to the Majorana bispinor field equation for a noninertial frame
referred to a curvilinear coordinate system: namely,
\begin{equation}
\fl\qquad
 V_\alpha^\mu(\mathrm x)\,\gamma^{\,\alpha}_{M}\left\{\partial_{\mu} - \mathrm i\x\Gamma_{\mu}(\mathrm x)\right\} \psi_M(\mathrm x)
+im\,\psi^{\,c}_M(\mathrm x)=0
\qquad\quad
\psi_M=\psi^{\,c}_M
\end{equation}


%
\end{document}